\documentclass[useAMS,usenatbib]{mn2e}

\usepackage{graphicx}
\usepackage{longtable}
\usepackage{lscape}
\usepackage{booktabs}
\usepackage{threeparttable}
\usepackage{subfigure}
\usepackage{graphics}
\usepackage{caption}
\usepackage{amsmath}
\usepackage{multicol}

\newcommand{\Garcia}{Garc{\'{\i}}a}

\title[Observations of HCN hyperfine line anomalies]
  {Observations of HCN hyperfine line anomalies towards low and high mass star-forming cores}
\author[|R. Loughnane et al.]
  {R.M.~Loughnane$^1$\thanks{E-mail: loughnane.robert@gmail.com (R. M. Loughnane); matt.redman@nuigalway.ie (M. P. Redman)},
  M.P.~Redman$^1$, M.A.~Thompson$^2$, N.~Lo$^{3,4,5}$, B.~O'Dwyer$^{1,6}$, 
  \newauthor M.R.~Cunningham$^4$\\
  $^1$Centre for Astronomy, School of Physics, National University of Ireland Galway, Ireland\\
  $^2$Centre for Astrophysics Research, Science and Technology Research Institute, University of Hertfordshire,\\ 
  College Lane, Hatfield, AL10 9AB, UK\\
  $^{3}$Departamento de Astronom\'ia, Universidad de Chile, Camino El Observatorio 1515, Las Condes, \\ Santiago, Casilla 36-D, Chile (current address)\\
$^{4}$School of Physics, University of New South Wales, Sydney, NSW 2052, Australia\\
$^{5}$Laboratoire AIM Paris-Saclay, CEA/Irfu - Uni. Paris Did\'erot - CNRS/INSU, 91191 Gif-sur-Yvette, France\\
  $^6$ Department of Applied Mathematics and Theoretical Physics, Wilberforce Road, Cambridge, CB3 0WA, UK\\
   }
\date{Submitted March 2011}

\pagerange{\pageref{firstpage}--\pageref{lastpage}} \pubyear{2010}

\def\LaTeX{L\kern-.36em\raise.3ex\hbox{a}\kern-.15em
    T\kern-.1667em\lower.7ex\hbox{E}\kern-.125emX}

\begin{document}
\label{firstpage}
\maketitle

\begin{abstract}
HCN is becoming a popular choice of molecule for studying star formation in both low and high mass regions and for other astrophysical sources from comets to high red shift galaxies. However, a major and often overlooked difficulty with HCN is that it can exhibit dramatic non-LTE behaviour in its hyperfine line structure. Individual hyperfine lines can be strongly boosted or suppressed. In low mass star forming cloud observations, this could possibly lead to large errors in the calculation of opacity and excitation temperature while in massive star forming clouds, where the hyperfine lines are partially blended due to turbulent broadening, errors will arise in infall measurements that are based on the separation of the peaks in a self-absorbed profile. This is because the underlying line shape cannot be known for certain if hyperfine anomalies are present. We present a first observational investigation of these anomalies across a wide range of conditions and transitions by carrying out a survey of low-mass starless cores (in Taurus and Ophiuchus) and high mass protostellar objects (in the G333 giant molecular cloud) using hydrogen cyanide (HCN) J=1$\rightarrow$0 and J=3$\rightarrow$2 emission lines. We quantify the degree of anomaly in these two rotational levels by considering ratios of individual hyperfine lines compared to LTE values. We find that {\it all} the cores observed demonstrate some degree of anomaly while many of the lines are severely anomalous. We conclude that HCN hyperfine anomalies are common in both lines in both low mass and high mass protostellar objects and we discuss the differing hypotheses for the generation of the anomalies. In light of the results, we favour a line overlap effect for the origins of the anomalies. We discuss the implications for the use of HCN as a dynamical tracer and suggest in particular that the J=1$\rightarrow$0, F=0$\rightarrow$1 hyperfine line should be avoided in quantitative calculations.
\end{abstract}


\section{Introduction}
While hundreds of molecules have now been detected in molecular clouds, there are only a handful of molecules that are useful as robust dynamical tracers.  A good probe species must be abundant enough to be readily observed and it must also be excited at high densities, so that it can trace the deep dense interior of molecular clouds (where the key dynamical processes take place) rather than just the low density outer layers. The species must also be chemically well behaved so that its abundance relative to hydrogen varies predictably. HCN matches all of these qualities and, at first sight, appears to be one of the very best tracers of molecular gas in space.
HCN was discovered in space by \citet{SnyBuhl71} and examples of the objects it has been recently used to observe include comets \citep{hogerheijde09,friedel05,hirota99}, planetary atmospheres \citep{marten02}, evolved star atmospheres \citep{schilke00,schilke03}, quiescent nearby low mass stellar nurseries \citep{yun99,park99}, distant massive star forming regions \citep{hennemann09,boonman01}, active galaxies \citep{kohno03} and molecular clouds in the high redshift universe \citep{gao04}. In particular, HCN has been used routinely as an infall tracer in low mass starless cores \citep{Taf06, Sohn07,choi99} and 
also in high mass star forming regions \citep{Wu05}.

HCN has a hyperfine structure due to the nuclear quadrupole moment of $^{14}{\rm N}$. This is potentially very useful since the optical depth and self-absorption could be measured by examining individual hyperfine lines while different rotational levels would give a measure of excitation temperature. 

\begin{figure}
\includegraphics[height=130pt]{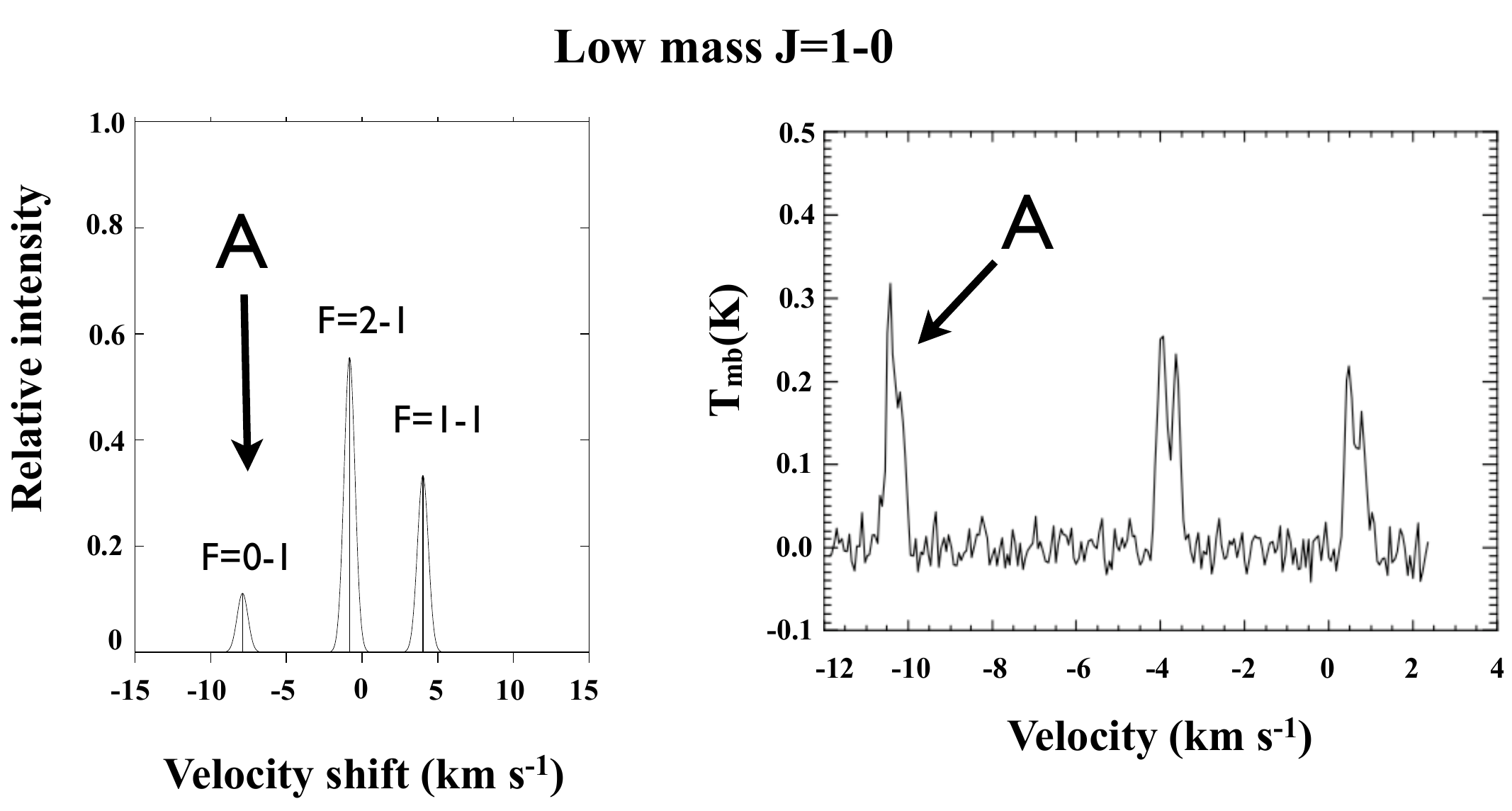}
\includegraphics[height=130pt]{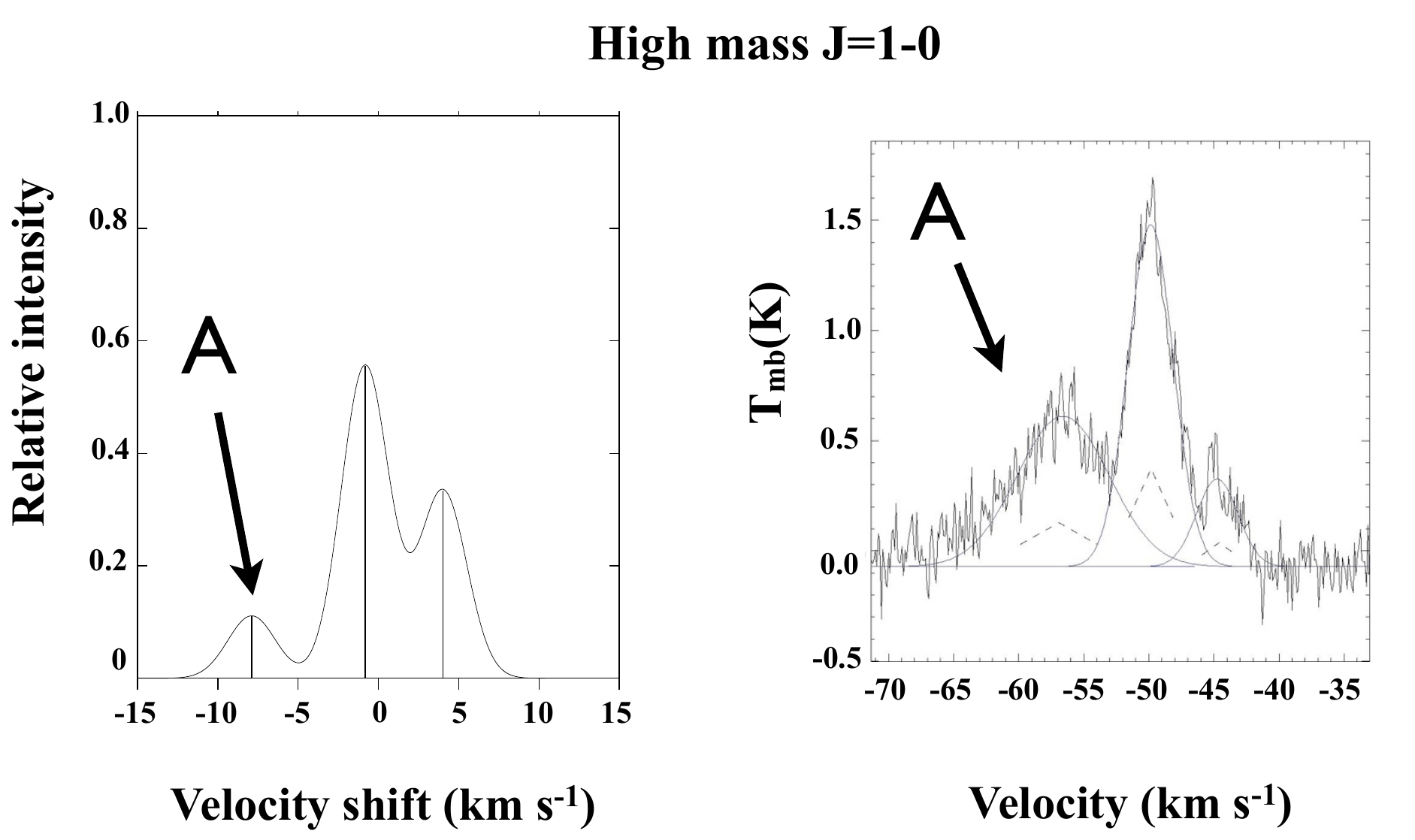}
\caption{Top: Expected line shape for optically thin HCN J=1$\rightarrow$0 in a low mass cold quiescent molecular cloud compared with a JCMT observation of a low mass protostellar core, L1197. The J=1$\rightarrow$0, F=0$\rightarrow$1 hyperfine line, marked 'A' is boosted far above its expected strength. Bottom: Expected line shape for optically thin HCN J=1$\rightarrow$0 in a massive turbulent molecular cloud (which increases the line width). This is compared with a Mopra observation of a massive core in the G333 cloud. Again the component marked `A' is boosted but is also much broader than the other components. This strongly suggests that line overlap effects at higher energy levels are significant.}
\label{lowmasshighmass10}
\end{figure} 
\begin{figure} 
\includegraphics[width=250pt]{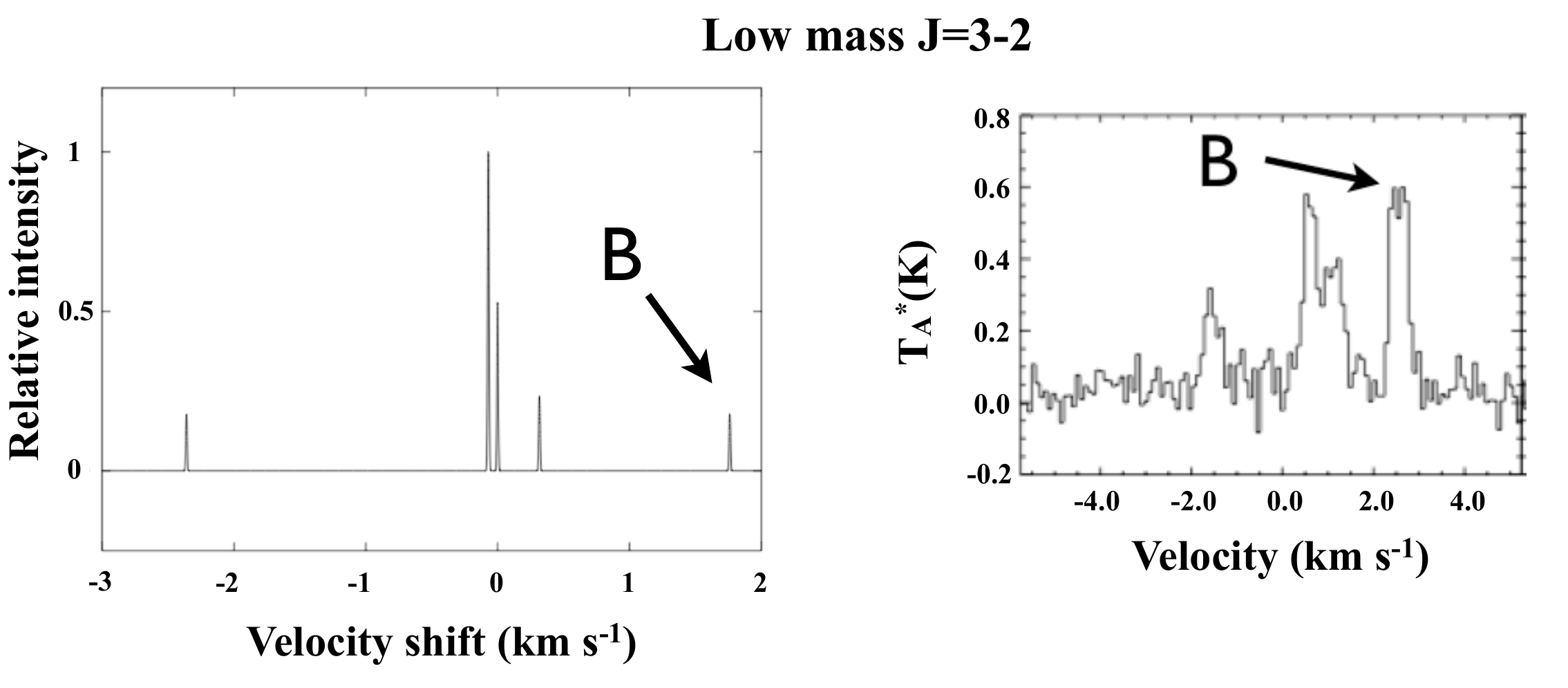}
\caption{The hyperfine structure of HCN J=3$\rightarrow$2 compared with a JCMT observation of L1622A2. The component marked 'B' is strongly anomalous. The laboratory spectroscopic rest frequencies and relative intensities of the hyperfine lines are listed in Table~\ref{tab:molspec}}
\label{lowmass32}
\end{figure} 
There is however, a major and often overlooked difficulty 
with using HCN for any quantitative calculations in that it has a rather pathological and puzzling hyperfine structure. In an early paper on this, Walmsley et al.\@~(1982)\nocite{Walms82}, using observations of HCN J=1$\rightarrow$0 in a low mass star forming cloud (TMC 1), demonstrated that the hyperfine components are present in ratios that mean they are not in thermal equilibrium with each other. The strengths of the individual hyperfine lines are `anomalous' in that they can appear boosted or suppressed far beyond what would be expected from an LTE analysis (or even from a standard non-LTE radiative transfer calculation). This phenomenon of hyperfine anomalies can also be seen in ${\rm N_2H^+}$  \citep{Keto10,Daniel07}, and deuterated nitrogen species \citep{turner01} {\bf but is not seen in other species with hyperfine lines, such as $\rm C^{17}$O, for example \citep{Redman02}}.  

The underlying mechanism for the HCN anomalies has never been fully settled with possible suggested causes including {\bf turbulent overlap, radiative scattering and line opacity effects (e.g. Guilloteau \& Baudry 1981; Cernicharo \& Guelin 1987; Gonzalez-Alfonso \& Cernicharo 1993; Turner et al 1997\nocite{Guill81,CernGue87,gonzalez93,turner97}). New observational data, coupled with radiative transfer codes that implement full line overlap offer the possibility of a solution to this long standing problem}, which will be timely in view of the beginning of the Atacama Large Millimeter Array (ALMA) era.

In this paper, an attempt is made to systematically investigate the HCN anomalies observationally. We present the combined results of a HCN molecular line survey carried out for this purpose toward 25 low mass starless cores in Taurus and Ophiuchus and towards 7 massive star forming turbulent cores in the G333 complex. We find that the anomalies are widespread in both classes of object. Our study marks the first identification in massive star forming regions of the same anomaly as observed in many of the low mass cores. We have in addition, for the first time, observed two anomalous rotational lines towards the same source. In \S2 we review the microphysics of the formation of the HCN hyperfine lines. \S3 details our source selection criteria, the molecular transitions and the telescopes used. In \S4, the results are presented and a method of systematically characterising the degree of anomalousness is described. An analysis of the hyperfine line ratios is then presented. Finally, in \S5, we summarise our findings and discuss the implications of our results.

\section{The HCN line profile}
HCN has a relatively high dipole moment ($\mu_{J=1\rightarrow0}$=2.98D\footnotemark\footnotetext{1 Debye [D] = 1$\rm0^{-18}$ statcoul cm} for 
HCN J=1$\rightarrow$0 versus $\mu_{J=1\rightarrow0}$=0.11D for CO J=1$\rightarrow$0) so that the lower transitions of HCN, 
especially, prove to be excellent tracers of dense molecular gas in star-forming clouds as well as in stellar 
complexes such as galaxies. This is due to the critical densities of rotational transitions obeying 
$\rm n_{crit}\propto\mu^{2}\nu^2_{J\rightarrow J-1}$, for optically thin lines at frequency 
$\rm \nu_{J\rightarrow J-1}$, allowing the HCN transition lines to trace $\approx$100-500 times denser 
gas than corresponding CO transitions \citep{Papa07}. HCN possesses a dominant end nitrogen atom that means it is less prone to surface effects on dust grain mantles, in particular freeze-out, compared to other more abundant $\rm H_2$ mass tracers such as HC$\rm O^{+}$ 
or $\rm C^{17}$O \citep{Freed05}. Thus HCN remains abundant in the gas phase in the cold central regions of star-forming cores.

\begin{table}
\begin{center}
\caption{Calculated Spectroscopic values for hyperfine components of the J=1$\rightarrow$0, 2$\rightarrow$1 and 3$\rightarrow$2 transitions (Aherns 2002)}
\begin{threeparttable}
\begin{tabular}{@{}ccrc@{}}
\hline \hline
&& Frequency &\\[-1ex]
\raisebox{1.5ex}{J F} & \raisebox{1.5ex}{$\rm J^{\prime} F^{\prime}$} & (GHz) & \raisebox{1.5ex}{S(hfs)}\\ [-0.5ex] 
\toprule
1 1 & 0 1 & 88.630413 & 0.3333\\
1 2 & 0 1 & 88.631846 & 0.5555\\
1 0 & 0 1 & 88.633935 & 0.1111\\
2 2 & 1 2 & 177.259676 & 0.0833\\
2 1 & 1 0 & 177.259921 & 0.1111\\
2 2 & 1 1 & 177.261109 & 0.2500\\
2 3 & 1 2 & 177.261220 & 0.4667\\
2 1 & 1 2 & 177.262010 & 0.0056\\
2 2 & 1 1 & 177.261109 & 0.0833\\
3 3 & 2 3 & 265.884887 & 0.0370\\
3 2 & 2 1 & 265.886185 & 0.2000\\
3 3 & 2 2 & 265.886431 & 0.2963\\
3 4 & 2 3 & 265.886497 & 0.4286\\
3 2 & 2 3 & 265.886976 & 0.0011\\
3 2 & 2 2 & 265.888519 & 0.0370\\

\bottomrule
\end{tabular}
\begin{tablenotes}
\item[] 
\end{tablenotes}
\end{threeparttable}
\label{tab:molspec}
\end{center}
\end{table}

The (nuclear quadrupole) hyperfine structure of HCN is solely due to the non-vanishing electric 
quadrupole moment of the end $\rm^{14}N$ nucleus ($\rm Q_{^{14}N}$=20.44 $\pm$ 0.03mb \footnotemark\footnotetext{mb $\Rightarrow$ milli-electron-barns, 1 barn=$\rm10^{-24}cm^2$}), which can lead 
to line splitting of the order of several MHz, resulting in an easily identified split line structure (with rotational transitions 
in the GHz range). By taking ratios of the relative intensities of neighbouring hyperfine lines one can constrain the optical depth of a 
given region in the cloud. Since hyperfine lines belonging to a particular rotational transition differ by a few MHz in frequency, there 
should be little concern regarding the coupling of the telescope beam. The physical properties of a source such as the 
density, the temperature, and molecular abundance depend on the optical depth and so the hyperfine structure should be taken into account 
upon analysing spectral line observations \citep{Keto10}.  

\footnotetext{mb $\Rightarrow$ milli-electron-barns, 1 barn=$\rm10^{-24}cm^2$}

Table~\ref{tab:molspec} gives the hyperfine line frequencies and their relative intensities.
Under LTE or optically thin conditions, the ratios of the relative intensities of the neighbouring hyperfine lines in each of the two rotational 
transitions considered here generally adhere to a set of fixed constants/limits. The relative weightings, given such conditions, in the case 
of the lower rotational transition are of the form 1:5:3 (or 0.111:0.555:0.333 in terms of relative intensities) for HCN  J=1$\rightarrow$0. For HCN  J=3$\rightarrow$2, four of the six hyperfine lines are not spectrally resolved and the spectrum takes the appearance of one strong central component and two satellite lines, with a ratio of 1:25:1 between them. 

As a preview of the observations presented below, Figures~\ref{lowmasshighmass10} and~\ref{lowmass32} show clear examples of the anomalies. Figure~\ref{lowmasshighmass10} shows the theoretically expected optically thin line profile for HCN  J=1$\rightarrow$0 for turbulent widths appropriate for low mass star formation regions (top left panel) and massive star formation regions (bottom left panel). Alongside these are example observations that clearly show the anomalies: the component marked A should go from being the weakest of the three hyperfine lines, in the optically thin limit, to equal in strength to the other components in the optically thick limit. Instead, in both types of region, it is either the broadest or strongest component of the profile. Figure~\ref{lowmass32} shows that a similar effect is present in the higher excitation HCN  J=3$\rightarrow$2 line.  As will be shown in this paper, it is clear that the hyperfine anomalies are present not only in the ground state of HCN in low mass star forming regions but also in other levels and in massive star forming environments. Since these anomalies are common this must currently render any quantative interpretation of the spectrum of HCN insecure at best. 

\section{Observations and Results}

\subsection{Low mass star forming clouds}
In order to quantify the presence of HCN anomalies in low mass star forming clouds, we collected observations of both the HCN J=1$\rightarrow$0 and HCN J=3$\rightarrow$2 lines for a selection of nearby low mass starless cores.

The sources were primarily chosen from a comprehensive HCN J=1$\rightarrow$0 survey
in HCN and HNC, catalogued by \citet{Sohn07} using the Taeduk Radio Astronomical Observatory (TRAO) in Korea. The majority of these objects, being well-known Taurus-Auriga and 
Ophiuchus sources, were selected by \cite{Sohn07} as being associated with the central peak of the (1,1) inversion transition of N$\rm H_3$, 
as observed by single-dish radio telescopes, and hence tracing dense gas \citep{Lee99}. The objects surveyed are found to be at various 
distances (see Table~\ref{tab:CoreLineParams}, at the end of the manuscript) and have luminosities in the range 0.1 to 0.5 $\rm L_{\sun}$. \citet{Sohn07} kindly supplied us with their data which form all of our HCN J=1$\rightarrow$0 line profiles.

\begin{table*}
\label{tab:Obsparams}
\begin{minipage}{110mm}
\caption{Observational parameters related to individual telescopes}
\begin{threeparttable}
\begin{tabular}{cccccccccc}
\hline \hline
& Frequency & Backend & $\Delta \nu$ & $\Delta v$ &&\\[-1ex]
\raisebox{1.5ex}{Transition} & (GHz) &  & (kHz) & (km$\rm s^{-1}$) & $\raisebox{1.5ex}{$\rm F_{eff}$}^\ddagger$ & \raisebox{1.5ex}{$\rm \eta_{mb}$} \\ [-0.5ex]
(1) & (2) & (3) & (4) & (5) & (6) & (7)\\ 
\toprule
J=1$\rightarrow$0 & 88.6339360 & TRAO auto & 10 & 0.033 & 0.58 & 0.50\\
J=3$\rightarrow$2 & 265.8864343 & JCMT ACSIS & 30.5 & 0.034 & 0.47 & 0.63\\ 
J=3$\rightarrow$2 & 265.8864200 & JCMT DAS & 78.1 & 0.088 & 0.47 & 0.63\\
J=3$\rightarrow$2 & 265.886487$\rm 0^{\dagger}$ & IRAM auto & 31 & 0.035 & 0.68 & 0.43\\
J=4$\rightarrow$3 & 354.505500 & KOSMA AOS & 53 & 0.059 & 0.46 & 0.64\\

\bottomrule
\end{tabular}
\begin{tablenotes}
\item[] \scriptsize{NOTES.-- (2) ${\dagger}$Frequency of the J,F = 3,4$\rightarrow$2,3 hyperfine component as given by the CDMS, http://www.ph1.uni-koeln.de/vorhersagen/. \\
(6) $\rm F_{eff}$ = $\Omega_{2\pi}$/$\Omega_{4π}$ and $\Omega_{2\pi}$($\Omega_{4\pi}$) is the integral of the beam pattern over
the forward hemisphere $4\pi$, \citep{Bensch01}.}
\end{tablenotes}
\end{threeparttable}
\end{minipage}
\end{table*}

The cores to be observed in HCN J=3$\rightarrow$2 were selected on the basis of how bright the central  F=2$\rightarrow$1 hyperfine component was in the HCN J=1$\rightarrow$0 rotational transition since it was anticipated that if this particular hyperfine transition was strong then 
the probability for a detection in the J=3$\rightarrow$2 rotational transition of HCN was improved. Cores that were also clearly anomalous in the HCN J=1$\rightarrow$0 line were further prioritised.

Table~2 summarises the telescopes and setups used to collect the data. The bulk of the HCN J=3$\rightarrow$2 (265.8862 GHz) emission line observations for the low mass sources were carried out at the James Clerk Maxwell Telescope (JCMT) between September 2007 and July 2008 in band 5 ($\tau_{225}\ge$0.2) conditions. Single-point observations were obtained for 30 sources in position-switching  mode with a pre-defined off-position of 50$\rm 0^{\prime\prime}$ in an arbitrary direction. The ACSIS digital autocorrelation spectrometer with a bandwidth of 250MHz was used providing  a velocity resolution of $\sim$0.034$\rm km~s^{-1}$. The receiver noise temperature (DSB mode) was 510-850K. The telescope main-beam efficiency was 0.69
and the half-power beam width (HPBW) was $\sim$2$\rm0^{\prime\prime}$ in the range 225-285GHz.  

Data from supplementary sources were gathered from the JCMT archive. This data consisted of 
HCN J=3$\rightarrow$2 line observations towards six starless cores L1495A-N, L1544, L1517B, L1512, L1521F and L1622A-2 which were carried out between February and March 
2005, typically in band 2 ($\tau_{225}\ge$0.06) conditions. The DAS (which preceded ACSIS at the JCMT) was used in frequency switching mode and with a velocity resolution of 0.088~${\rm km~s^{-1}}$. The system temperature during the course of this observing run varied between 250 and 380K. These starless cores were mapped in HCN J=3$\rightarrow$2, in a strip of five successive positions across each core. We used the central position data in the analysis below but data such as these could be used to investigate how the anomalies vary within a single source.

Additional data, from IRAM and KOSMA, for two sources complete our database. These are J=3$\rightarrow$2 and J=4$\rightarrow$3 data for the two starless cores, 
L1498 and TMC-1 respectively, both part of the Taurus-Auriga complex. The former data has been extracted from a molecular survey of two prototypical 
starless cores carried out by \citet{Taf06} whilst the latter core data was supplied from line observations taken by \citet{Ahrens02}. 
Table~\ref{tab:CoreLineParams}, at the end of the manuscript, lists all the sources observed and gives the detections statistics. There were 11 sources with no detection. 

Figure~3 displays HCN J=1$\rightarrow$0 and J=3$\rightarrow$2 line profiles for most of the low mass sources. For the low mass sources it can be seen that the HCN J=1$\rightarrow$0 hyperfine lines are well separated and resolved.  By reference to Figures 1 and 2 it is readily apparent that the HCN anomalies are present in many of the sources; the leftmost hyperfine line should never exceed in strength either of the other two hyperfines. For many of the sources, the individual line profiles exhibit either red or blue asymmetry. Such signatures are interpreted as being due to dynamical activity such as infall, outflow, rotation or pulsation that gives doppler shifts in excess of the thermal/turbulent line widths (e.g. \citealt{mardones97,carolan08,Redman04,alves01,Redman06}).  Examining these dynamical properties is beyond the scope of this work but many of the low mass sources have been widely investigated elsewhere. However, it is very much worth pointing out that for several of these sources, the HCN J=1$\rightarrow$0, F=0$\rightarrow$1 (highest frequency, blue-wards) hyperfine has the reverse of the red or blue asymmetry present in the other two hyperfines (e.g. L1521B, L204C-2, L234E-S). This is not likely to be explicable in terms of the bulk dynamics taking place in the source (see, for example, Stahler \& Yen 2010 \nocite{stahler10} who model such asymmetries in the HCN hyperfine lines, but do not address the anomalies) and is instead likely to be part of the microphysics of the hyperfine anomalies. This effect is returned to in the discussion section.  

As described in \S2 the low mass HCN J=3$\rightarrow$2 profile can be considered to be composed of a strong central component with two satellite lines. Again, the anomalies are readily apparent in many of the sources. The left satellite line is heavily suppressed and the right satellite is boosted to comparable or higher strength than the central component. The central component is marginally resolved at the thermal and turbulent line widths present in low mass star forming regions so the line shape is complex as a result of partial blending and dynamical effects in the source. The right satellite would provide a better guide to the underlying line shape of each hyperfine in these conditions. For the sources with detections, Table \ref{tab:HCNcoreprops}, at the end of manuscript, gives the antenna temperature of the strongest hyperfine and the integrated HCN J=1$\rightarrow$0 and HCN J=3$\rightarrow$2 hyperfine line intensities. 

\subsection{Massive star forming regions}
An excellent example of a massive star forming region is the G333 molecular cloud associated with the RCW 106 H{\sc ii} region. This giant molecular cloud complex spans  an $\sim 0.7~{\rm deg}^2$  region, and is $\sim 3.6~{\rm kpc}$ away. G333 has been the subject of a large multi-molecular line legacy survey using the 22-m Mopra Telescope of the Australia Telescope National Facility (ATNF) between 2004 and September 2006 (see e.g. \citep{Lo09,wong08,bains06}).  The data presented in this paper were collected with the 8-GHz wide band Mopra spectrometer (MOPS) centred at 87 GHz. The FWHM beamsize is $\approx$36$^{\prime\prime}$ \citep{Ladd05}. Full details of the observational setup is available in \citet{Lo09}. 
\begin{table*}
\begin{minipage}{\textwidth}
\begin{center}
\begin{threeparttable}
\caption{HCN J=1$\rightarrow$0 Observations towards G333 cores}
\begin{tabular}{ccccc}
\hline \hline
Emission & R.A. & Dec. & $\rm T_A^*$ & $\rm V_{lsr}$ \\[-0.2ex]
\raisebox{0.4ex}{Peak} & (J2000) & (J2000) & (K) & (km$\rm s^{-1}$)\\[-0.4ex]
(1) & (2) & (3) & (4) & (5)\\
\toprule
G332.741-0.619 (MMS73) & 16 20 07.0 & -50 59 50 & 0.864 & -50.28\\
G332.772-0.589 (MMS71) & 16 20 05.7 & -50 57 04 & 1.372 & -55.94\\
G332.813-0.703 (MMS75) & 16 20 49.7 & -51 00 15 & 0.755 & -53.61\\ 
G333.096-0.503 (MMS50) & 16 21 14.2 & -50 39 44 & 1.861 & -55.57\\
G333.241-0.517 (MMS38) & 16 21 52.5 & -50 34 13 & 0.740 & -49.09\\  
G333.293-0.423 (MMS30) & 16 21 42.9 & -50 28 09 & 1.440 & -50.18\\
G333.297-0.357 (MMS26) & 16 21 25.6 & -50 24 45 & 1.527 & -49.82\\ 

\bottomrule
\end{tabular}
\begin{tablenotes}
\item[] \footnotesize{NOTES.-- (1) - Name of source using its galactic coordinates and Mookerjea et al (2004)\nocite{mookerjea04} identifier; (2), (3) - Right Ascension (R.A.) and Declination (Dec.) of each source in their corresponding equatorial coordinates; (4) Antenna temperature of the central F=2$\rightarrow$1 hyperfine component; (5) Local standard of rest (LSR) velocity of central hyperfine component.} 
\end{tablenotes}
\end{threeparttable}
\end{center}
\end{minipage}
\label{tab:HiMassObs}
\end{table*}

We used this Mopra data set to identify individual massive star forming cores in HCN J=1$\rightarrow$0 emission. Because this region is distant and very confused, and multiplicity is common in massive star formation, it was important for the present study to identify clear isolated examples of massive cores. HNC is often co-spatial to HCN but is less abundant (though see Sarrasin et al 2010\nocite{sarrasin10}) and exhibits no hyperfine structure\footnotemark. Therefore, clumps of HNC emission that exhibit single peaked line profiles are likely signposts of isolated dense cores that will be bright in HCN. We identified seven new HCN sources in the G333 data cube in this way. These sources were detected in the 1.2 mm emission catalogue G333 of Mookerjea et al (2004) \nocite{mookerjea04} and their cold temperatures and dust masses ranging from $\sim 400 - 1200~{\rm M_\odot}$, mark them as cold massive cores.  The positions of the seven sources are displayed in Figure~4 and are noted in Table~3.

\footnotetext{HNC does actually possess an underlying hyperfine structure, but the location of the $\rm^{14}N$-nucleus, means that the individual components are unresolvable and a single peak is observed}

\begin{figure*}
\begin{center}
\includegraphics[scale=0.6]{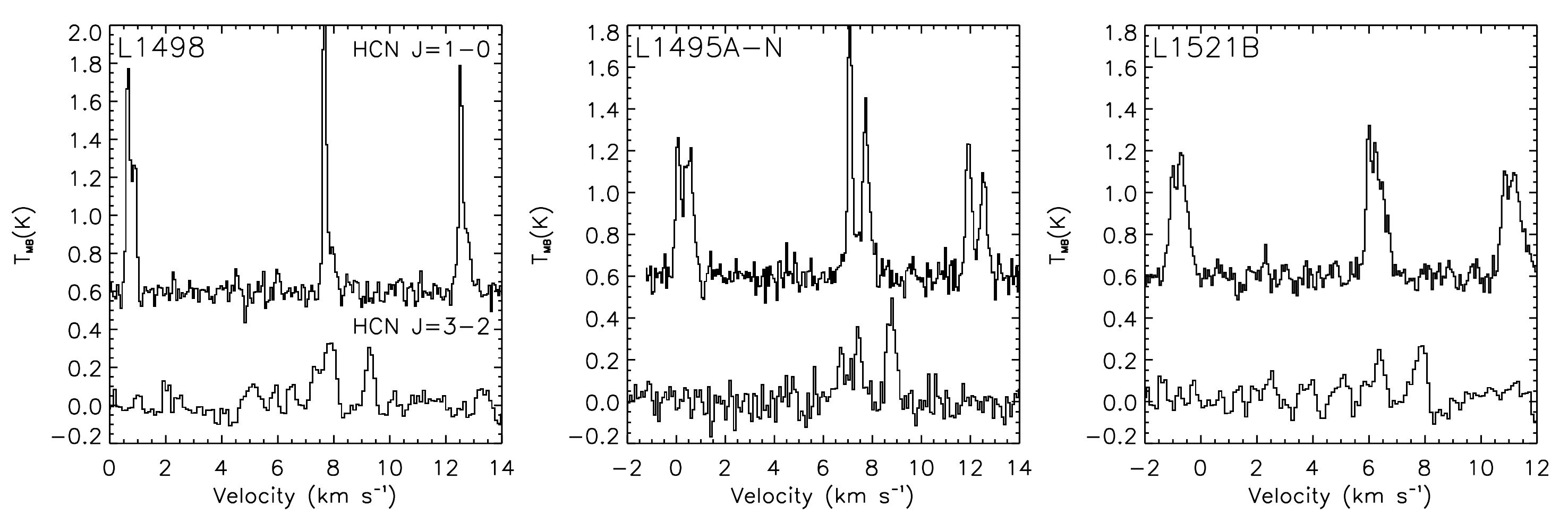}\\		
\includegraphics[scale=0.6]{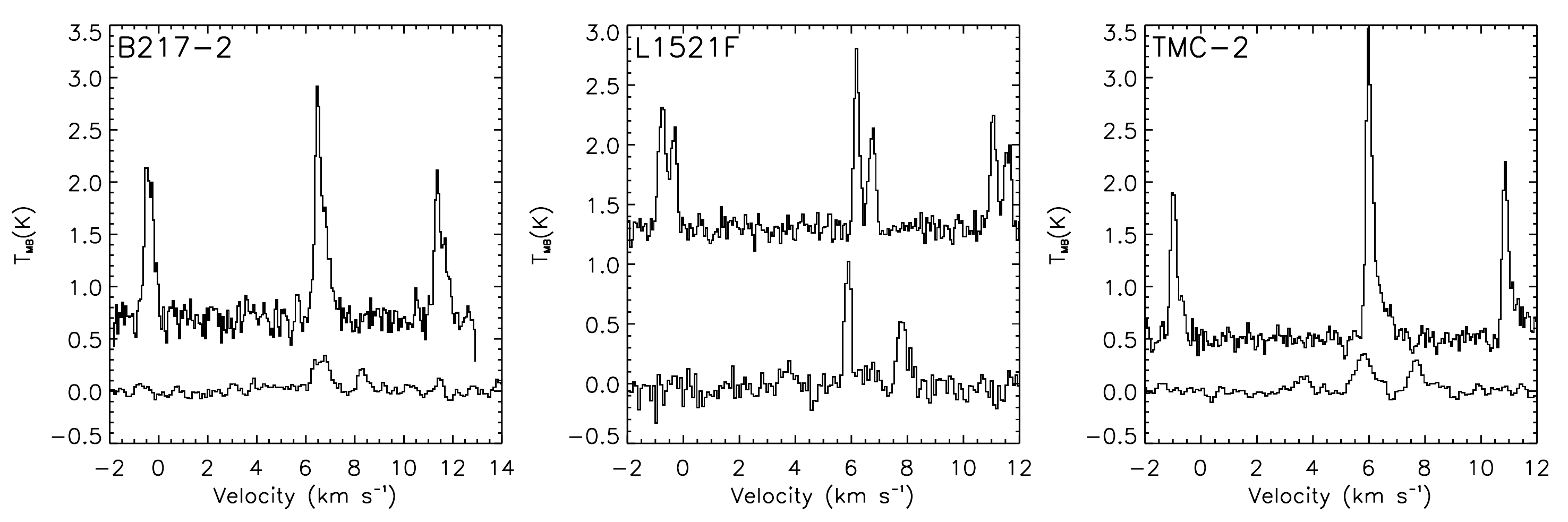}\\
\includegraphics[scale=0.6]{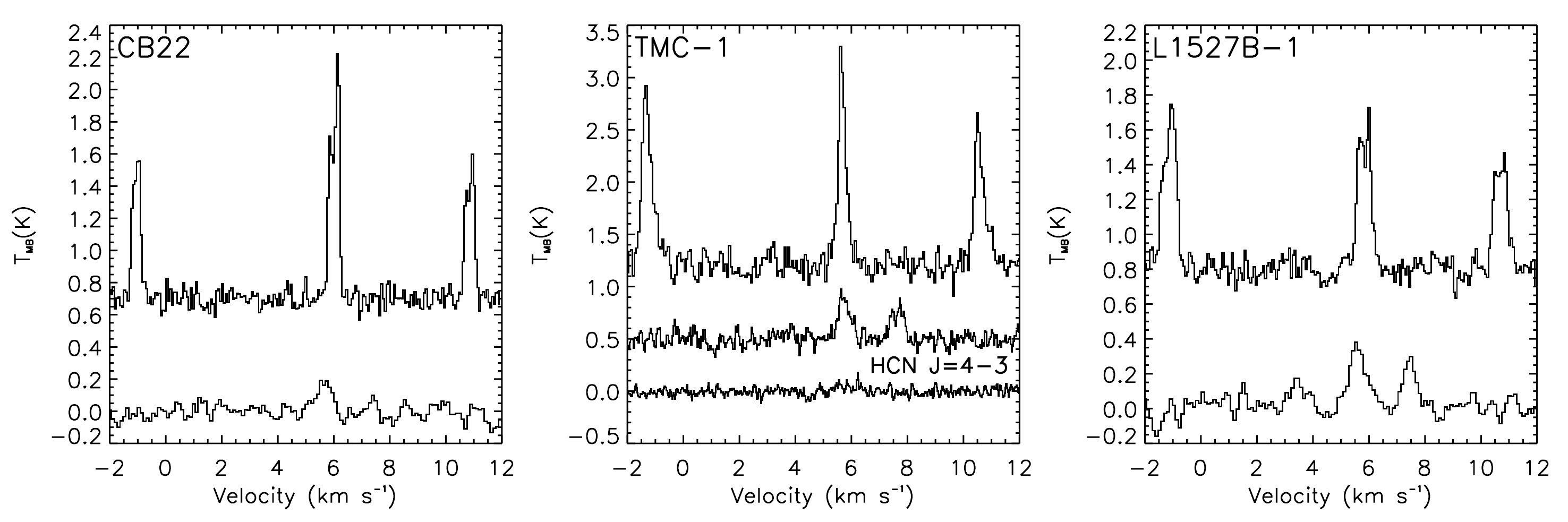}\\
\includegraphics[scale=0.6]{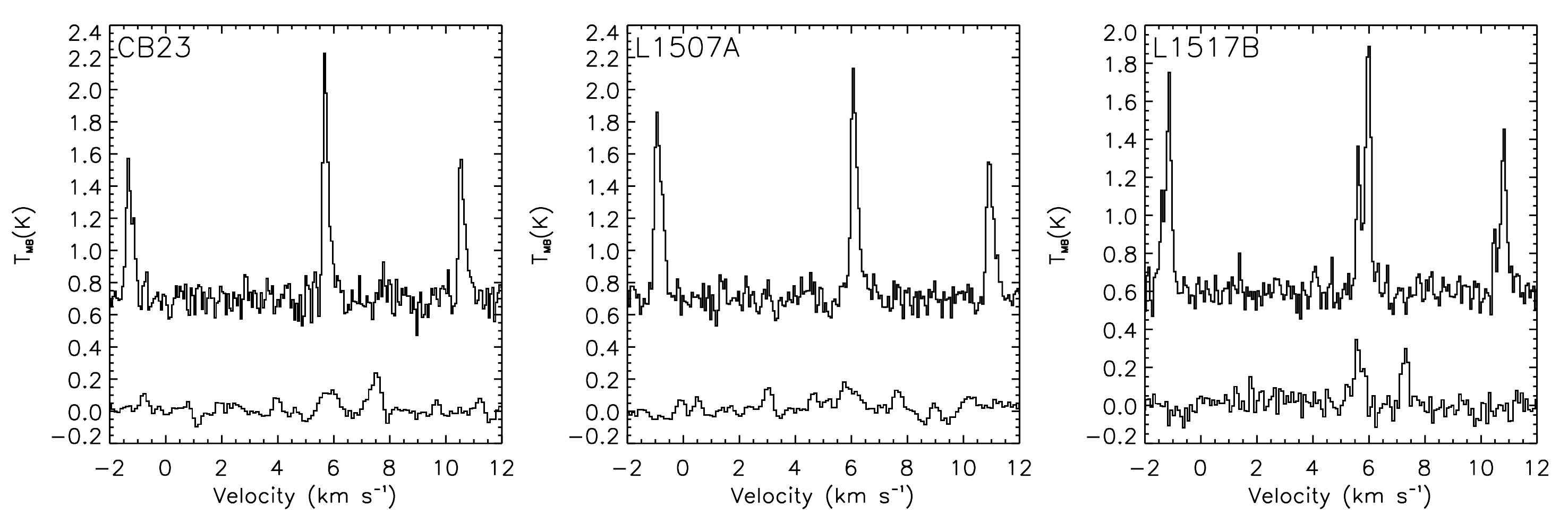}
\end{center}
\caption[ContinuedFloat]{Respective HCN (i) J=1$\rightarrow$0 and (ii) J=3$\rightarrow$2 towards 29 low mass starless cores with a 
different degree of anomaly present in each transition towards a specific source of emission.}
\label{CoreSpectra1}
\end{figure*}

\begin{figure*}
\begin{center}
\includegraphics[scale=0.6]{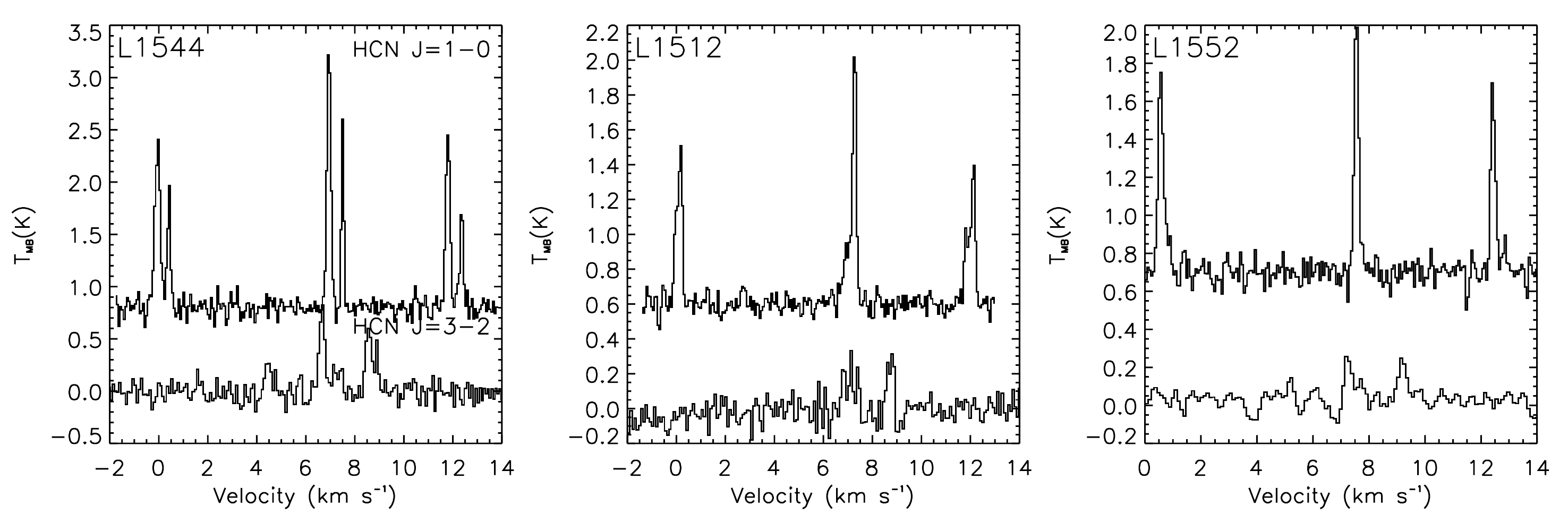}\\		
\includegraphics[scale=0.6]{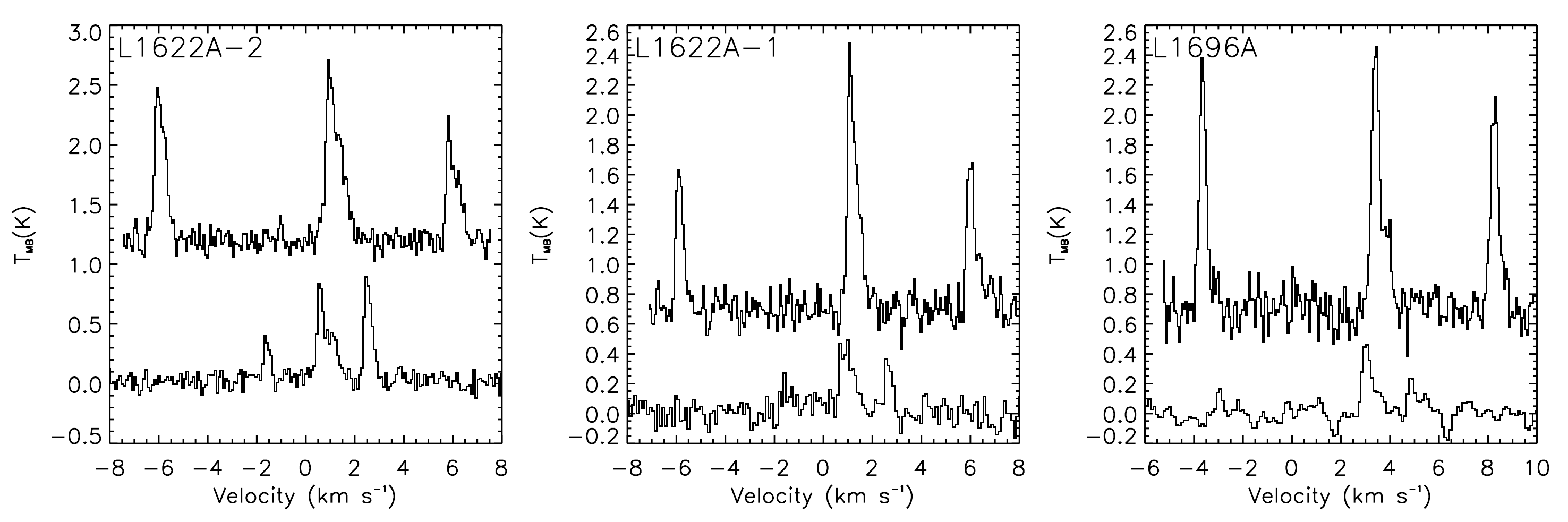}\\
\includegraphics[scale=0.6]{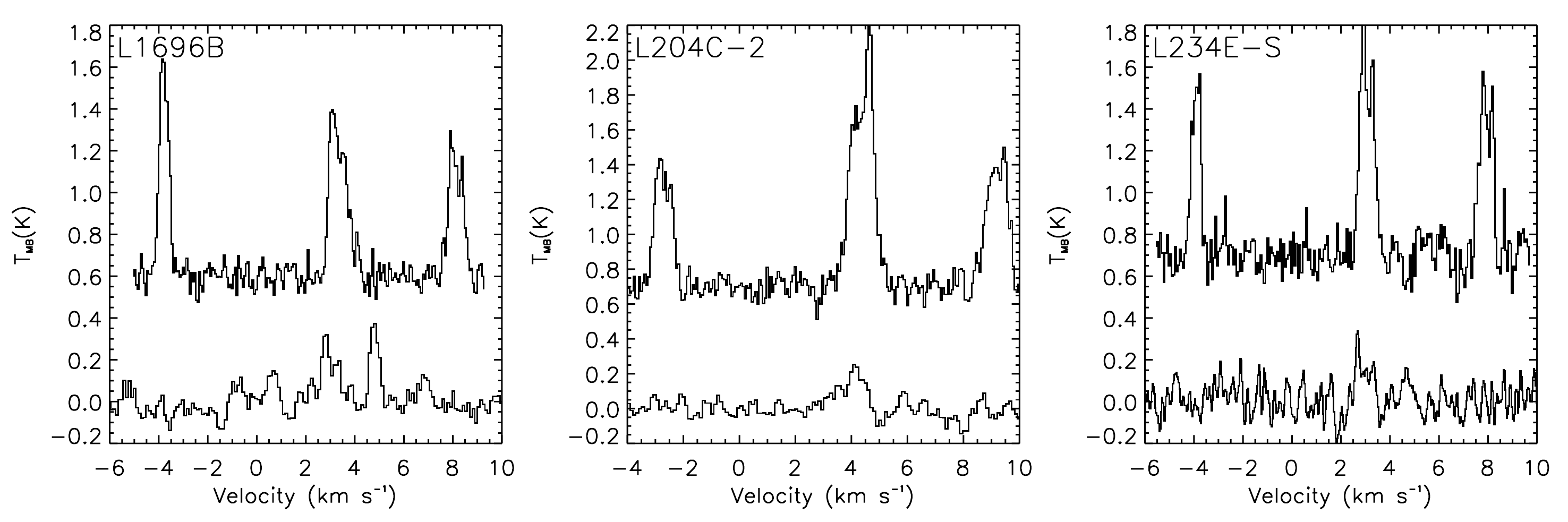}\\
\includegraphics[scale=0.6]{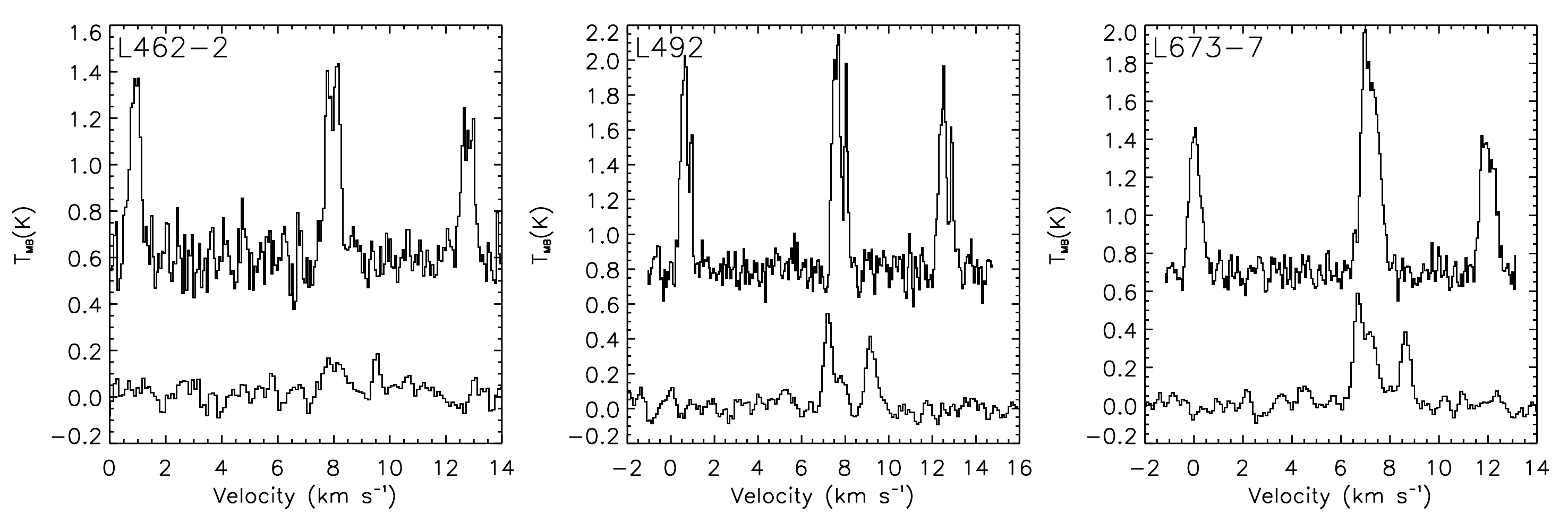}
\end{center}
{\bf Figure 3 \it contd.}
\label{CoreSpectra2}
\end{figure*}

\begin{figure*}
\begin{center}
\includegraphics[scale=0.6]{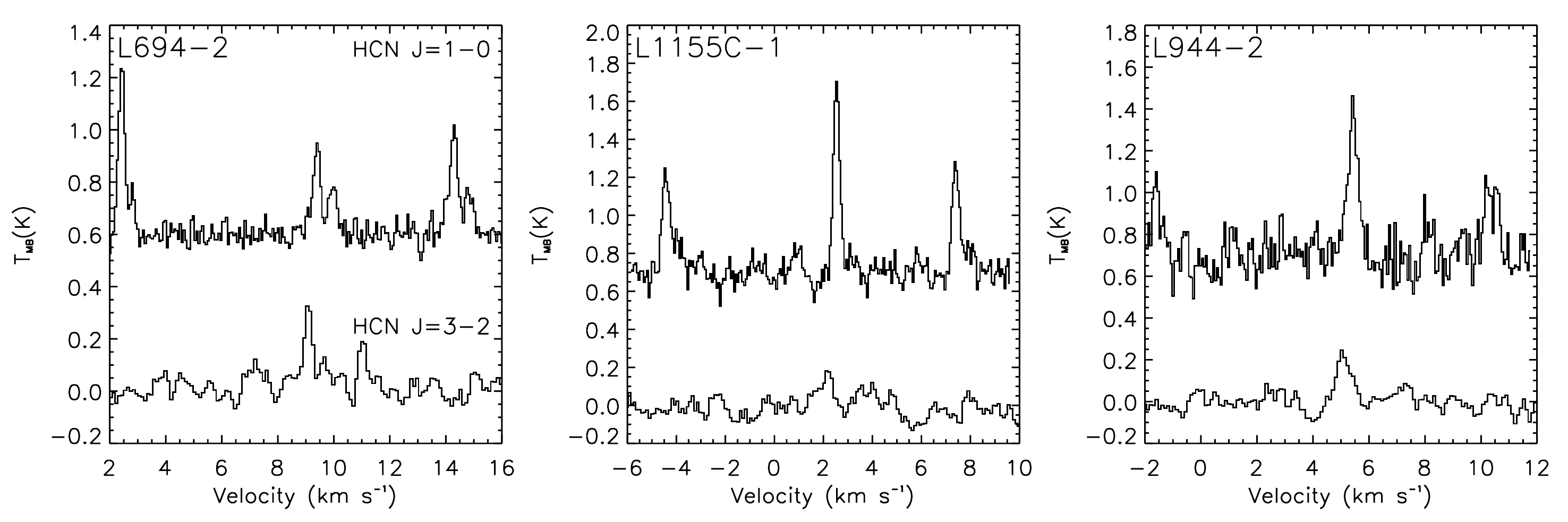}\\
\includegraphics[scale=0.6]{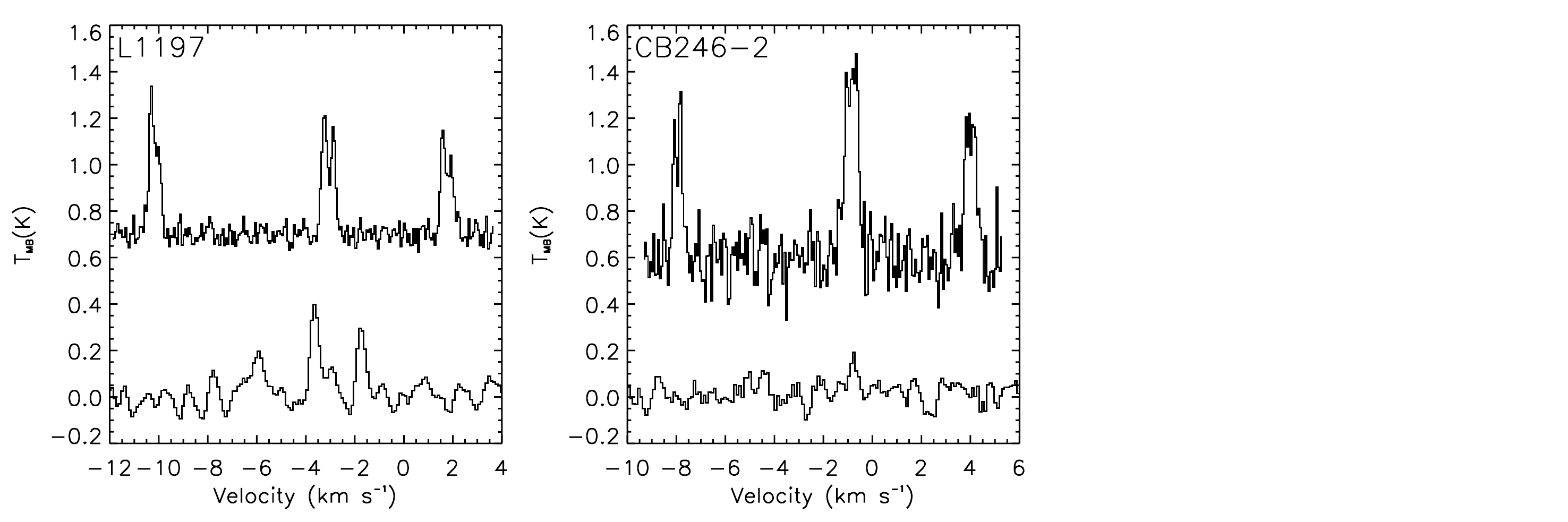}
{\bf Figure 3 \it contd.}
\end{center}
\end{figure*}

\begin{figure*}
\begin{center}
\includegraphics[scale=0.4]{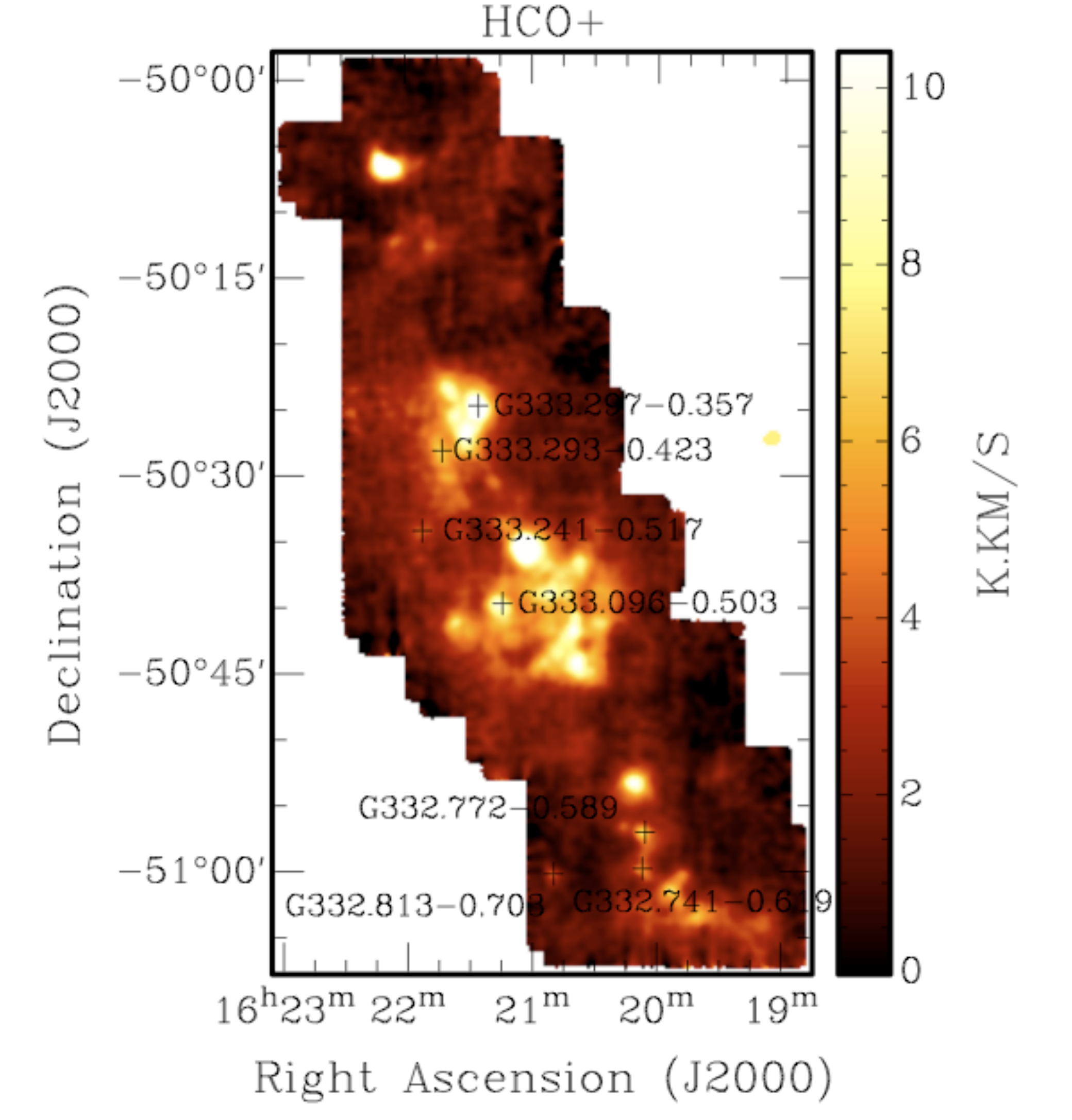}
\includegraphics[scale=0.4]{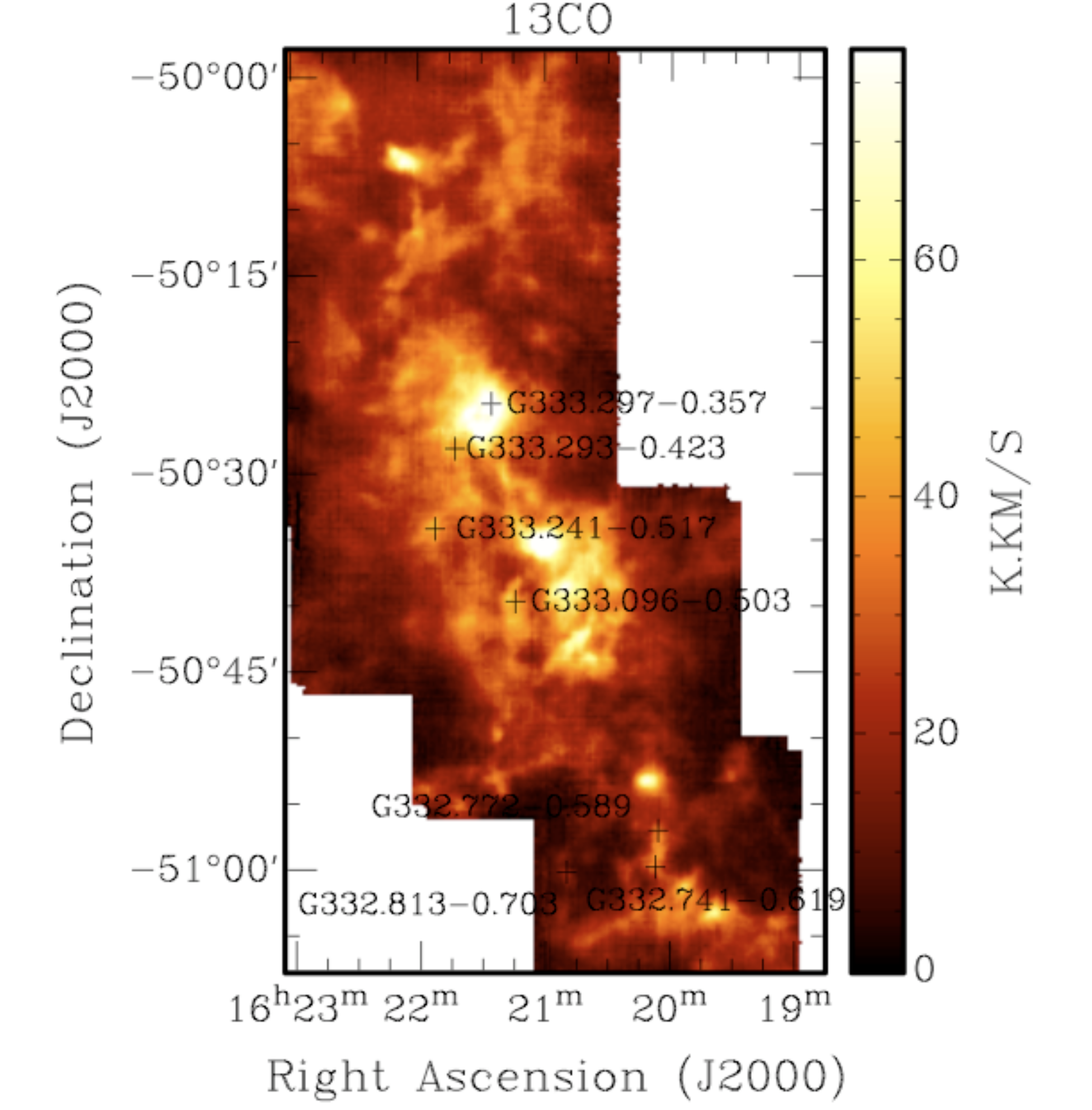}
\caption{HCO$^+$ J=1$\rightarrow$0 (left) and $^{13}$CO J=1$\rightarrow$0 (right) integrated intensity (zeroth-moment) maps of the G333 giant molecular cloud. The crosses mark the positions of the seven HCN anomaly emission sources presented in this work.}
\end{center}
\label{fig:G333cores}
\end{figure*}

\begin{figure*}
\includegraphics[scale=0.8]{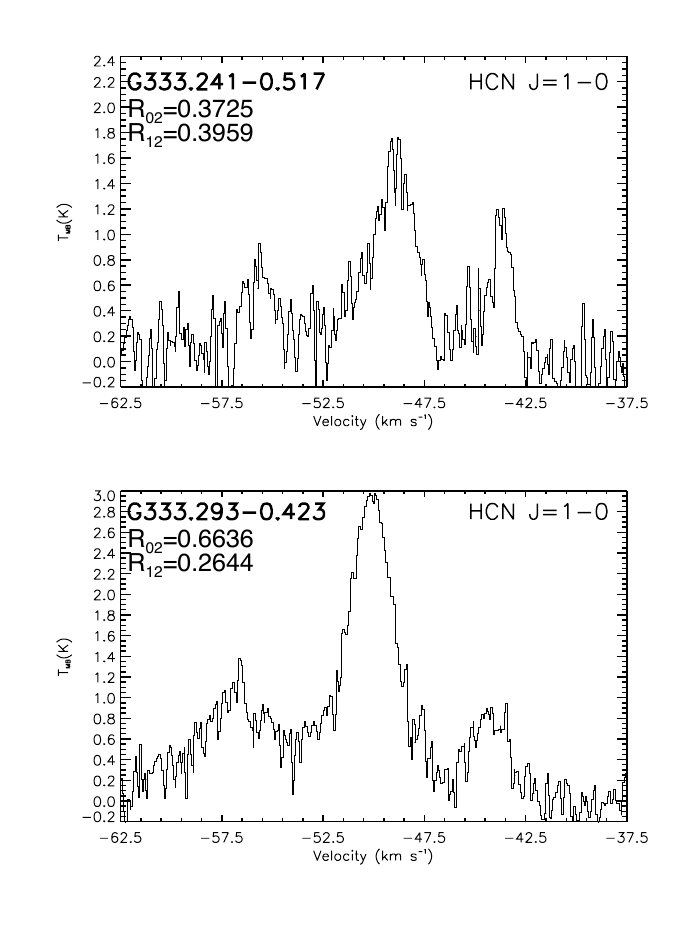}
\includegraphics[scale=0.8]{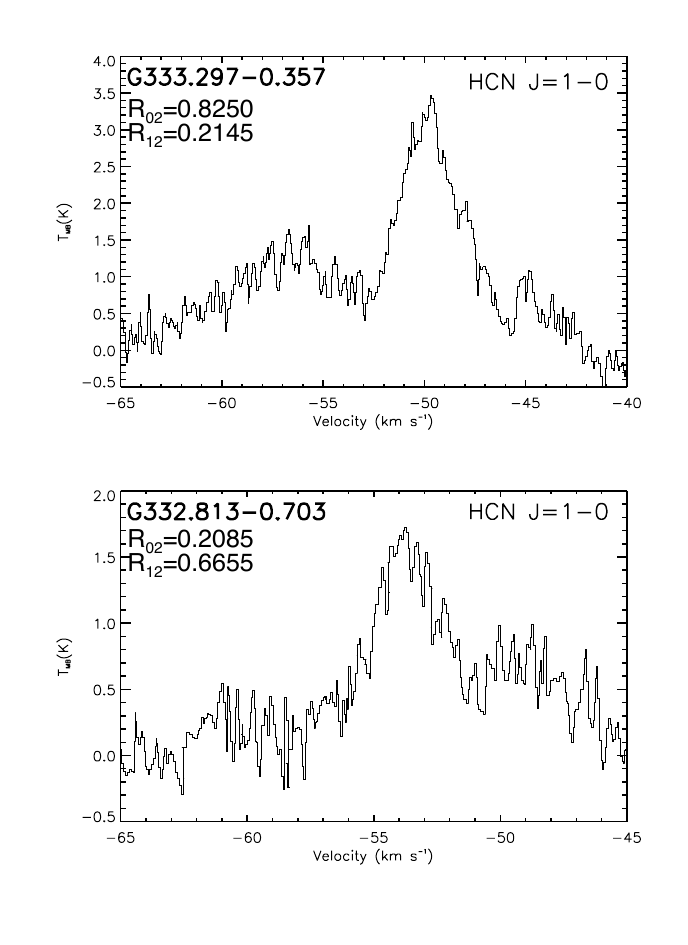}
\includegraphics[scale=0.8]{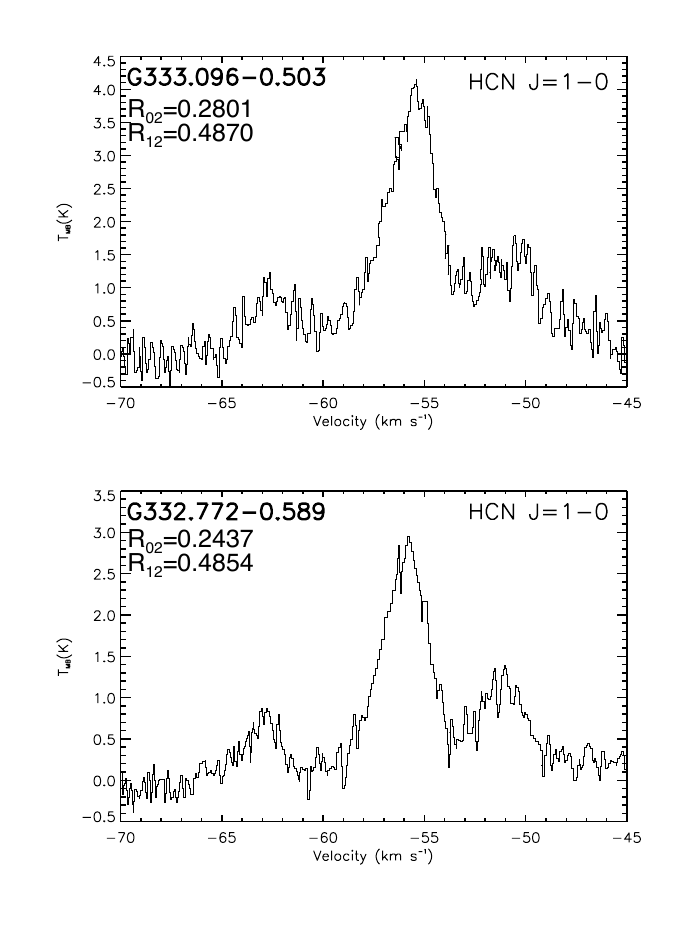}
\includegraphics[scale=0.8]{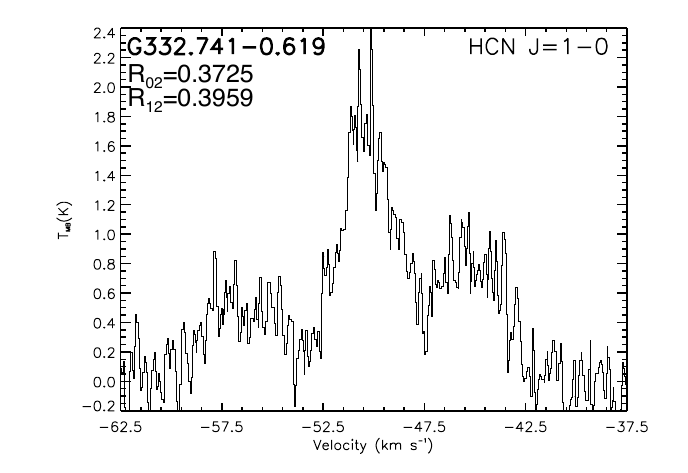}
\caption{HCN J=1$\rightarrow$0 line profiles from the seven 
G333 GMC sources.}
\label{fig:G333profiles}
\end{figure*}

Figure~\ref{fig:G333profiles} displays the HCN J=1$\rightarrow$0 line profiles from these seven cores.  For these massive core HCN J=1$\rightarrow$0 lines, the three individual hyperfine components are still clearly identifiable but, due to the greater degree of turbulence, the line widths are much greater and are comparable to the separation of the hyperfines. The resulting hyperfine line parameters for the HCN J=1$\rightarrow$0 data for each of the seven massive cores are summarised in Table~3. The anomalies here are of a notably different character in that now the {\it line widths} of the individual hyperfines can be different as well as the line strengths. The presence of a variation in the line widths of individual hyperfine lines within a single rotational transition is an important result in the context of the origin of the anomalies and is returned to in the discussion section later in the paper.

\section{Hyperfine intensity analysis}

Reviewing Figures~1 and~2 and the full catalogue of line profiles in Figures~3 and~5, the anomalies are readily apparent visually. In order to characterise the anomalies in a quantative fashion, we here define the methods used to quantify the degree of anomalousness in the two rotational transitions. The character of the anomalies is then analysed for the two types of star forming region.

\subsection{HCN J=1$\rightarrow$0}
For the HCN J=1$\rightarrow$0 transitions, we use the method first utilised by \citet{Gott75} where the ratios of the individual hyperfine component integrated intensities, I, are measured and compared with Local Thermodynamic Equilibrium (LTE) values. These ratios, R(I[F=1$\rightarrow$1]/I[F=2$\rightarrow$1]) and R(I[F=0$\rightarrow$1]/I[F=2$\rightarrow$1]) (hereafter, simply $\rm R_{12}$ and $\rm R_{02}$) were calculated for each core and compared with the ranges expected for non-anomalous values. 

The ratios $R_{02}$ and $R_{12}$ equal 0.2 and 0.6, respectively, under LTE or optically thin conditions. Under optically thick conditions, the lines tend to saturation such that they are of equal intensity i.e. $R_{02}=R_{01}=1$. Thus the expected, non-anomalous, ranges are 0.2$<$$\rm R_{02}$$<$1 and 0.6$<$$\rm R_{12}$$<$1. Cores 
with hyperfine spectra displaying ratios lying outside each of the two ranges are described as being the most anomalous, with those outside only 
one of the above ranges being intermediary anomalous candidates. Those sources of emission with ratios lying within the two ranges represent the non-anomalous cores. The integrated HCN J=1$\rightarrow$0 line intensity ratios are recorded in Table 4.

\subsection{HCN J=3$\rightarrow$2}
There are six possible hyperfine lines in HCN J=3$\rightarrow$2 but due to blending (the hyperfine splitting decreases strongly with rotational level) it is not possible to resolve the central group of hyperfine lines and the spectrum takes the appearance of one strong central component and two satellite lines.  The central group of three hyperfines can be identified as a collective branch that result from a net change of 1 in the total angular momentum quantum number, $\rm\Delta F$=1. One hyperfine branch also belonging to this central group is negligibly weak. The remaining two branches each result in an isolated hyperfine line. With this blending, the optically thin form of the intensity weightings is 1:25:1 (or 0.037:[0.200+0.296+0.428]:0.037 $\Rightarrow$ 0.037:0.925:0.037).

It is practical then to measure the ratio of the outlying hyperfines to the central component in order to quantify the degree of anomalousness. We introduce the expressions $\rm R_{\rm\Delta F(0^{-}\rightarrow1)}$ and $\rm R_{\rm\Delta F(0^{+}\rightarrow1)}$, hereafter $\rm R_{0^{-}1}$ and $\rm R_{0^{+}1}$ which represent the ratios of the central hyperfine branch ($\rm\Delta F=1$) compared with that of leftmost ($\rm\Delta F=0^{-}$) and rightmost ($\rm\Delta F=0^{+}$) hyperfine branches respectively. The integrated HCN J=3$\rightarrow$2 line intensity ratios are recorded in Table 4. 

\subsection{Anomalies in low mass star forming regions}
Table 4 summarises our analysis of the collected datasets of the two transitions towards the twenty-nine low mass starless cores that were detected in both lines. Examination of the table reveals that while the anomalies are common, the degree of anomalousness varies between sources. For example, the lower transition ratios are $\rm R_{02}$=0.95 and $\rm R_{12}$=0.80 for  L1197, and $\rm R_{02}$=1.18  and $\rm R_{12}$=1.08 for L1544 (recall that a value of 1 in the case of either ratio represents the LTE condition of saturation in each of the respective emission lines). In the higher transition, the values of the ratios are $\rm R_{0^{-}1}$ = 0.28 (six times in excess of the LTE value) and $\rm R_{0^{+}1}$ = 0.78 in the case of L1197 while for L1544 the ratios are $\rm R_{0^{-}1}$ = 0.64 (fifteen times in excess of the expected LTE value) 
and $\rm R_{0^{+}1}$ = 0.58. Thus it can be seen that L1544 is more anomalous than its Ophiuchus counterpart. 

\begin{table*}
\begin{minipage}{\textwidth}
\begin{center}
\begin{threeparttable}
\caption{Hyperfine intensity ratios in the HCN J=1$\rightarrow$0 and J=3$\rightarrow$2 transitions assembled for each of the 29 detected cores in the 
higher transition.}
\begin{tabular}{lcccc}
\hline \hline
 & \multicolumn{2}{c}{J=1$\rightarrow$0} & \multicolumn{2}{c}{J=3$\rightarrow$2}\\[-1ex]
\raisebox{1.5ex}{\textsc{SOURCE}} & $\rm R_{02}$ & $\rm R_{12}$ & $\rm R_{0^-1}^{a}$ & $\rm R_{0^+1}^{a}$ \\ [+0.5ex] 
\toprule
L1498 & 0.9771 & 0.8400 & 0.1214 & 0.7214 \\
L1495AN & 0.7148 & 0.6033 & 0.1212 & 0.8303 \\
L1521B & 0.8259 & 0.8750 & 0.1786 & 1.0952 \\ 
B217-2 & 0.6788 & 0.6182 & 0.2570 & 0.4134 \\
L1521F & 0.9189 & 0.8007 & 0.1818 & 0.6126 \\
TMC-2 & 0.4517 & 0.6409 & 0.3081 & 0.6860 \\
CB22 & 0.5792 & 0.6062 & 0.1667 & 0.1806 \\
TMC-1 & 1.0986 & 0.8563 & 0.2212 & 0.7212 \\
L1527B-1 & 1.0833 & 0.7500 & 0.3568 & 0.6000 \\
CB23 & 0.6162 & 0.6263 & 0.0408 & 1.3878 \\
L1507A & 0.8482 & 0.6205 & 0.3071 & 0.3228 \\
L1517B & 0.7046 & 0.5232 & 0.1000 & 0.5889 \\
L1544 & 1.1822 & 1.0659 & 0.2800 & 0.7822 \\
L1512 & 0.6337 & 0.6188 & 0.0602 & 0.6626 \\
L1552 & 1.0146 & 0.8540 & 0.6118 & 1.0471 \\
L1622A2 & 0.6774 & 0.4899 & 0.6571 & 0.2408 \\
L1622A1 & 0.7931 & 0.6207 & 0.3333 & 0.4815 \\
L1696A & 0.6982 & 0.6802 & 0.1771 & 0.5029 \\
L1696B & 0.7516 & 0.6553 & 0.2727 & 0.6753 \\
L204C-2 & 0.4156 & 0.5497 & 0.0065 & 0.0323 \\
L234E-S & 0.5335 & 0.7230 & 0.0543 & 0.4565 \\
L462-2 & 0.7227 & 1.1992 & 0.1845 & 0.5631 \\
L492 & 0.7266 & 0.7448 & 0.2933 & 0.6356 \\
L673-7 & 0.4082 & 0.5082 & 0.1061 & 0.4364 \\
L694-2 & 1.1368 & 1.1263 & 0.2922 & 0.3312 \\
L1155C1 & 0.7418 & 0.7198 & 0.0822 & 0.6575 \\
L944-2 & 0.4606 & 0.5394 & 0.2683 & 0.4024 \\
L1197 & 0.9470 & 0.7955 & 0.6443 & 0.5772 \\
CB246-2 & 0.4713 & 0.6245 & 0.3387 & 0.3065\\
\bottomrule
\end{tabular}
\begin{tablenotes}
\item[] \scriptsize{$\rm^{a}T$he ratios $\rm R_{0^{-}1}$ and $\rm R_{0^{+}1}$ are defined in relation to the J=3$\rightarrow$2 hyperfine structural branches in \S4.2.}
\end{tablenotes}
\end{threeparttable}
\end{center}
\end{minipage}
\label{tab:varanoms}
\end{table*}

Figure 6 displays $\rm N_2H^+$ J=1$\rightarrow$0 column density calculated from optical depth data from \citet{crapsi05}, who observed the majority of our sources, against the $R_{02}$ (top) and $R_{12}$ (bottom) ratios. Recall that in LTE these both of these line ratios must be in the range $0.2<R_{02}<1$ and $0.6 < R_{12}<1$. Each of the hyperfine intensity ratios, $\rm R_{02}$ and $\rm R_{12}$ demonstrate a steady increase as a function of the column density of an optically thin tracing species ($\rm N_2H^+ J=1\rightarrow0$). 
This result is largely in keeping with earlier predictions \citep{Zin87}. The correlation coefficients calculated from the plotted points are $r=0.5282$ for $R_{02}$ and $r=0.5216$ for $R_{12}$. We consider these to be strong correlation coefficients and therefore conclude that the denser the optically thin line along the line of sight, the more anomalous the corresponding HCN J=1$\rightarrow$0 spectrum and that, from this data, it appears that it is the $\rm R_{12}$ hyperfine ratio that is most responsive to the $\rm N_2H^+$ J=1$\rightarrow$0 column density.

\subsection{Anomalies in massive star forming regions}
As noted above, a crucial finding in our G333 massive cores is that in all cases one or both of the two outlying HCN J=1$\rightarrow$0 hyperfines are significantly broader than the central hyperfine. For example, the line widths of the three hyperfine components of G333.297-0.357 are 7.8, 4.2 and $3.4~{\rm km~s^{-1}}$, a variation of a factor of more than two.  As a result of this, to quantify the anomalies, we use the integrated line intensity rather than just the line strength. This method, similar to that used by \citet{Lapinov89} is described in the Appendix. The hyperfine intensity ratios towards high mass cores were formally computed on the basis of the ratio of their 
respective line strengths, e.g. $\rm R_{01}$=$\rm T_A^*(F=0\rightarrow1)/T_A^*(F=2\rightarrow1)$. This approach has been applied to the IR 
source NGC 7538 IRS1 by \citet{1993Cao}, where the respective hyperfine components were subject to considerable partial convergence on account 
of different broadening schemes and microturbulence. The results of the analysis of the G333 HCN J=1$\rightarrow$0 data are presented in Table 5 which demonstrates that while the majority of these cores are anomalous in this transition, the line ratios do not exceed unity. 
 
 \begin{table*}
\begin{minipage}{\textwidth}
\begin{center}
\begin{threeparttable}
\caption{Summary of HCN J=1$\rightarrow$0 Hyperfine Component Analysis towards G333 Cores}
\begin{tabular}{cccccccc}
\hline \hline
Emission & &  $\Delta v$ & $\int T_A^*$dv\\[-1ex]
\raisebox{0.5ex}{Peak} & \raisebox{1.5ex}{$\rm F\rightarrow F^{\prime}$} & (km$\rm s^{-1}$) & (K~km$\rm s^{-1}$) \\ [-0.7ex] 
(1) & (2) & (3) & (4)\\ 
\toprule
G332.741-0.619 &  0$\rightarrow$1 & 4.983 & 1.572$\pm$0.093\\
&                 2$\rightarrow$1 & 3.685 & 3.389$\pm$0.122\\
&	          1$\rightarrow$1 & 4.573 & 2.161$\pm$0.101\\
G332.772-0.589 &  0$\rightarrow$1 & 2.654 & 1.030$\pm$0.070\\
&		  2$\rightarrow$1 & 2.894 & 4.229$\pm$0.078\\		
&		  1$\rightarrow$1 & 3.140 & 2.052$\pm$0.077\\
G332.813-0.703 &  0$\rightarrow$1 & 4.057 & 0.627$\pm$0.096\\
&                 2$\rightarrow$1 & 3.746 & 3.009$\pm$0.084\\
&		  1$\rightarrow$1 & 5.203 & 2.002$\pm$0.092\\
G333.096-0.503 &  0$\rightarrow$1 & 3.836 & 1.879$\pm$0.121\\
&		  2$\rightarrow$1 & 3.387 & 6.709$\pm$0.147\\
&		  1$\rightarrow$1 & 3.967 & 3.267$\pm$0.142\\
G333.241-0.517 &  0$\rightarrow$1 & 2.654 & 0.926$\pm$0.107\\
&	          2$\rightarrow$1 & 2.894 & 2.486$\pm$0.101\\
&		  1$\rightarrow$1 & 3.140 & 0.984$\pm$0.119\\
G333.293-0.423 &  0$\rightarrow$1 & 5.879 & 3.239$\pm$0.091\\
&		  2$\rightarrow$1 & 3.185 & 4.881$\pm$0.107\\
&		  1$\rightarrow$1 & 2.612 & 1.290$\pm$0.071\\
G333.297-0.357 &  0$\rightarrow$1 & 7.824 & 5.707$\pm$0.109\\
&		  2$\rightarrow$1 & 4.256 & 6.917$\pm$0.111\\
&	          1$\rightarrow$1 & 3.452 & 1.484$\pm$0.106\\

\bottomrule
\end{tabular}
\begin{tablenotes}
\item[] \footnotesize{NOTES. -- (1) - Source name in galactic coordinates; (2) - Respective hyperfine component; (3) - Line width of each hyperfine component in km$\rm s^{-1}$; (4) - Integrated intensity of individual hyperfine components in ${\rm K~km~s^{-1}}$}.  
\end{tablenotes}
\end{threeparttable}
\end{center}
\end{minipage}
\label{tab:HiMassProps}
\end{table*}
 
We do not have corresponding HCN J=3$\rightarrow$2 data for the G333 sources but line profiles have been obtained in this transition towards some massive star forming regions. Examples can be seen in Wu \& Evans (2003), Wu et al (2005) and Carolan et al (2009). At first glance, it would be expected that the hyperfine structure is not resolved in these sources. However, Carolan et al (2009) found that in order to correctly model the HCN J=3$\rightarrow$2 spectra, the underlying hyperfine line structure needed to be taken into account in order to correctly measure dynamical processes such as infall. Regarding any hyperfine anomalies, the heavy blending would render an analysis similar to that undertaken for the HCN J=3$\rightarrow$2 low mass source data very difficult if not impossible.

\begin{figure*} 
\includegraphics[width=400pt]{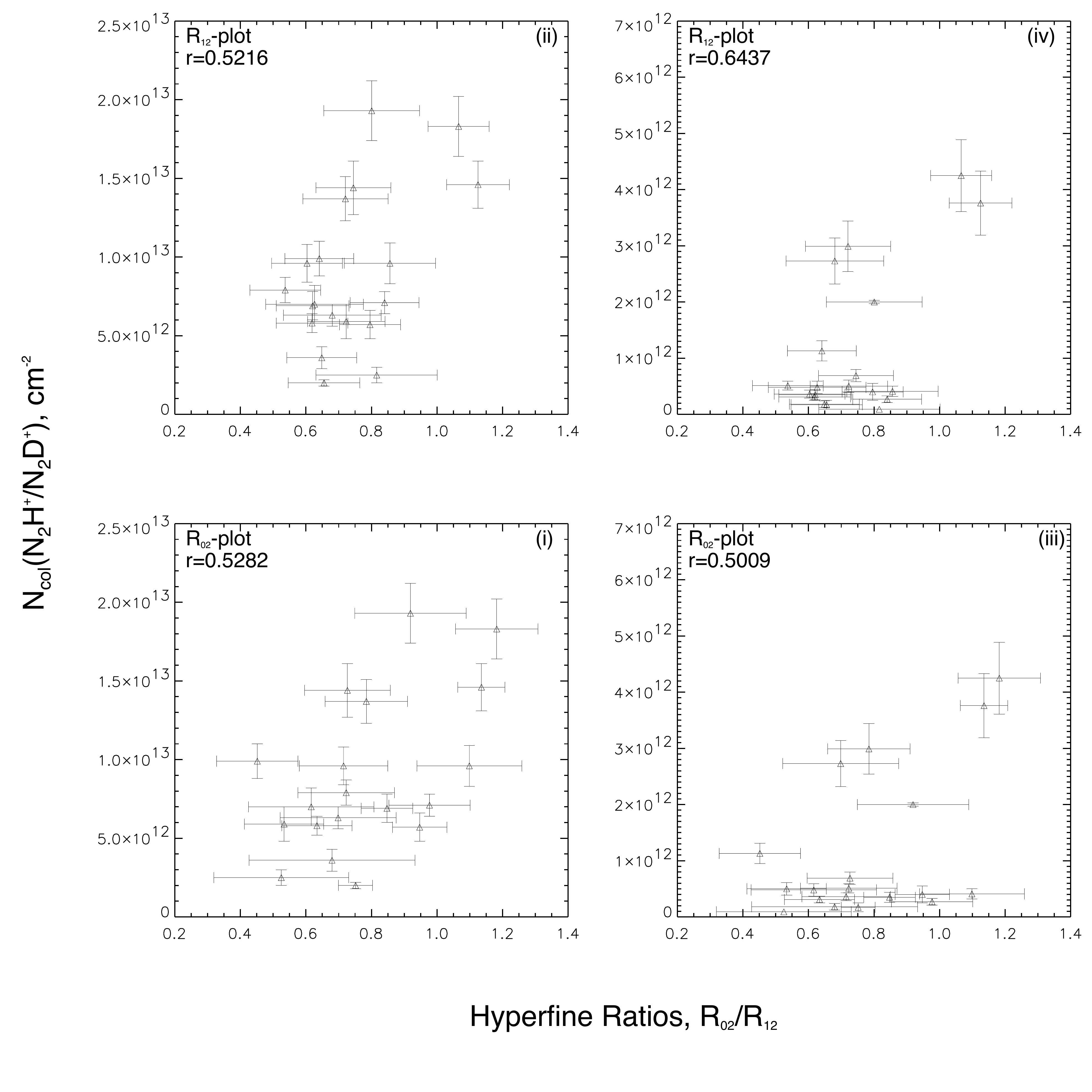}
\caption{Hyperfine ratios $R_{02}$ (top) and $R_{12}$ (bottom) versus ${{\rm N}_2{\rm H}^+} 1 \rightarrow$0 column densities. The correlation coefficient $r$ is recorded in the top left corner.}
\label{n2hdensvratios}
\end{figure*} 

\subsection{Characterising the anomalies}
\begin{figure*}
\begin{center}
\subfigure{%
\label{fig:lowmassrat}
\includegraphics[scale=0.4]{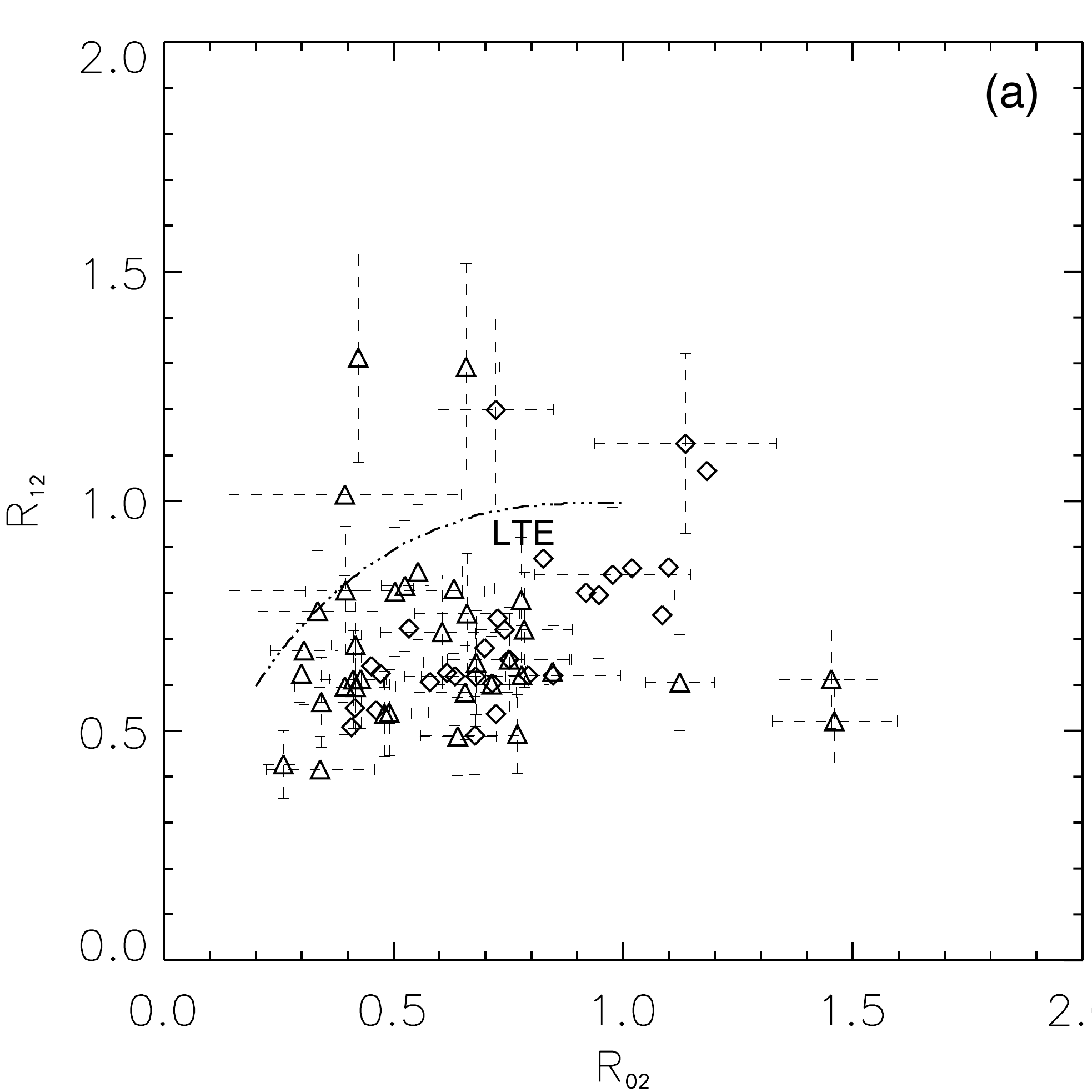}}
\hspace{0.5cm}
\subfigure{%
\label{fig:highmassrat}
\includegraphics[scale=0.4]{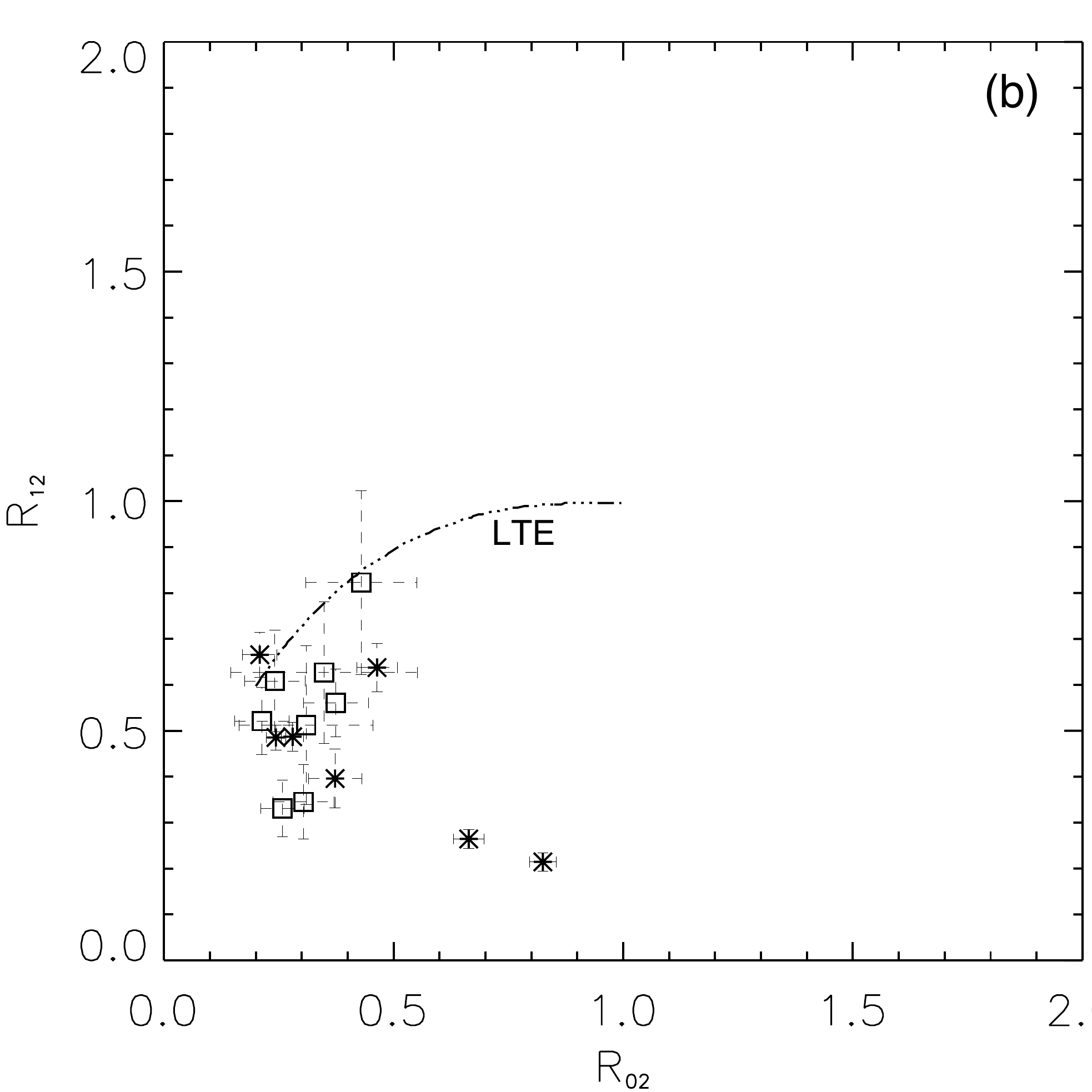}}
\subfigure{%
\label{fig:compmassrat}
\includegraphics[scale=0.4]{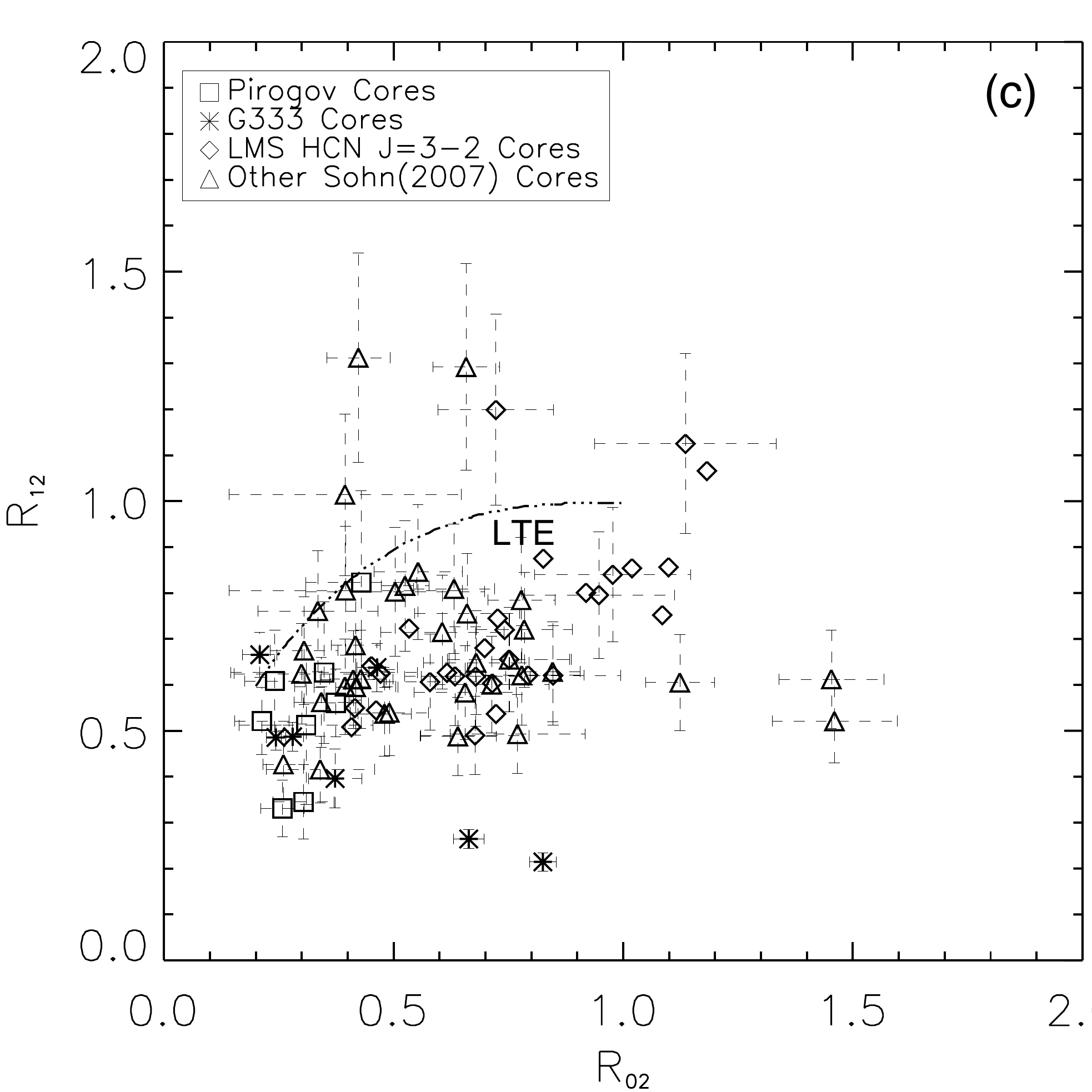}}
\end{center}
\caption{HCN J=1$\rightarrow$0 hyperfine ratios, $\rm R_{12}$ plotted against $\rm R_{02}$, for (a) the 64 low-mass starless sources in Table 1 of \citet{Sohn07} and (b) the 15 high-mass objects considered in this work, made up from the seven G333 cores indicated in Fig.~\ref{fig:G333cores} and the eight sources in figure~1 of \citet{Pirog99}. In (c), both sets of object's ratios are plotted in tandem. LTE ratios for different optical thicknesses would lie on the dashed curves in each figure.}
\label{fig:ratiosgraph}
\end{figure*}

In Figure~\ref{fig:ratiosgraph}, we plot the $\rm R_{02}$ ratio against the $\rm R_{12}$ ratio for both sets of cores. For the low-mass cores, this involved plotting the ratios after calculations using the integrated intensity values in Table~\ref{tab:HCNcoreprops}, with the respective errors along each dimension also computed. In relation to the higher mass cores, the G333 objects were plotted using the ratios calculated from the values indicated Table~5. In the high-mass plot, Fig.~\ref{fig:highmassrat}(b), we also include the corresponding hyperfine intensity ratios for the eight sample HCN J=1$\rightarrow$0 spectra for the sources indicated in figure~1 of \citet{Pirog99}. For both sets of readings, we computed corresponding errors based on the analysis using the technique outlined in Appendix A. It can be seen from comparing Figs.~\ref{fig:lowmassrat}(a) and \ref{fig:highmassrat}(b) that the ratios in the high mass cores are more tightly constrained with respect to their collective hyperfine ratios than the low mass sources. 
This, together with the fact that the high mass cores considered as part of this study do not display hyperfine ratios exceeding unity, indicate that warmer regions tend to form anomalous HCN J=1$\rightarrow$0 spectra that favour the pumping of the central F=2$\rightarrow$1 hyperfine component, that is, the thermal overlap description considered in \citet{Guill81}. This explanation will be returned to in \S5. Due to the relative low densities of our candidate low-mass objects, these figures may also reveal that the conditions conducive to such regions favour pathways of excitation at the quantum level that differ from the LTE routes, as discussed in \S5.

\section{Discussion}

\subsection{Hypotheses for the origin of HCN hyperfine anomalies}
The HCN hyperfine anomalies are unlikely to be due to a simple non-LTE opacity effect because i) the anomalies are seen in regions of varying column density and non-anomalous clouds are seen with a similar range of density \citep{Lapinov89} and ii) non-LTE radiative transfer calculations do not reproduce the anomalies \citep{stahler10}. Several hypotheses have been presented in the last few decades to account for the anomalies. Radiative trapping in the hyperfine transitions \citep{Kwan75}, can be ruled out upon closer examination of the molecular physics involved \citep{Lapinov89}. {\bf Collisional rates for HCN for conditions in dark clouds are now well established (see Sarrasin et al.\@ 2010 for the latest rates, noting also Faure et al.\@ 2007) \nocite{faure.et.al07,sarrasin10} so poor rate data are unlikely to be the cause of the anomalies.} \citet{Gott75} developed an idea which attempted to address the problem by way of probable de-excited photons from the hyperfine lines of upper rotational transitions. This introduced the possibility of line overlap for the first time. \citet{Guill81} developed on this theory in their thermal overlap treatment of the J=2$\rightarrow$1 transition for clouds where T=30K. Their treatment demonstrated an overpopulation of the J=1,F=2 level at the expense of the J=1,F=1 and F=0 levels, respectively. With regard to the J=1$\rightarrow$0 rotational transition, such an emphasis in population leads to a strengthening of the  F=2$\rightarrow$1 main hyperfine line and a corresponding weakening the two side hyperfine components. However, our and other observations towards low mass cores show a relative strengthening in these side components so a simple thermal overlap effect involving a single level can be discounted.

\subsection{A line overlap effect for the HCN hyperfine anomalies}
The above studies did demonstrate that it is possible to change the strength of individual hyperfine lines through line overlap and this was further investigated by Zinchencko \& Pirogov (1987). A line overlap treatment involving several rotational levels and including line broadening due to a combination of thermal, doppler and turbulent effects could then offer a promising possible explanation for the anomalies. There are several pieces of evidence from our study that supports such an explanation. The lowest HCN transition gives rise to lines that are widely spaced in frequency. Transitions between higher rotational levels lead to reduced separations between the hyperfine lines, leading to blending beyond around the HCN J=4$\rightarrow$3 level in low mass clouds and the HCN J=3$\rightarrow$2 level in high mass clouds. The hyperfine lines arising from transitions between the J=4,3 and 2 levels could then be sensitive to doppler effects arising from, for example, enhanced or suppressed stimulated emission due to locally doppler shifted photons. This is perhaps what is been seen in two effects noted earlier in the paper. Firstly, in our low mass star forming clouds, several sources exhibited a mixture of blue and red asymmetric line profiles in the same rotational level transition but in all three cases (and noting the very small number statistics) it was the leftmost line (J=1$\rightarrow$0, F=0$\rightarrow$1) that differed from the other two. Secondly, in our high mass star forming clouds, the hyperfine line widths varied within the same rotational level transition. It is the leftmost hyperfine line involved again, this time being typically significantly wider than the other lines.  A line overlap in a higher rotational level transition could be the cause of these effects. In the very narrow lines in the low mass cloud, a small change in line strength at certain wavelengths could disturb the line shape. In the broad lines in the high mass cloud, the leftmost transition appears to have grown in width at the expense of the central component in particular. 

{\bf In the line overlap hypothesis, the anomalies increase with optical depth, which may also explain the reason why an isotopomer of HCN, H$^{13}$CN, does not exhibit anomalous intensity ratios \citep[e.g. for TMC-1][]{Irvine84}. This species, which is chemically identical to HCN and with an identical dipole moment, is lower in abundance by a factor of 40-50 and so is therefore typically optically thin. The apparent absence of an anomalous feature in this tracer�s hyperfine spectrum and its close relation to the more abundant HCN, indicates a role opacity effects must play in the formation of anomalous intensities. 

A limitation of our observational approach is that for the majority of sources, we were constrained to a single pointing towards the centre of the source. Cernicharo et al (1984) \nocite{Cerni84} presented the first large-scale map of the HCN J=1$\rightarrow$0  anomalies, in TMC-1. The anomalies were seen to be spatially extended but at the low opacity envelope at the edge of the source, the central F=2$\rightarrow$1 component remains strong relative to the two side components \citep{Guill81,CernGue87} while closer to the centre, the F=0$\rightarrow$1 component is at its strongest.  This could be another manifestation of the opacity dependence for the anomalies discussed above.  Cernicharo et al (1984) describe a concept by which radiative scattering of the photons emitted from the high density core is taking place at the lower density periphery. This mechanism can reproduce the extremely small ${\rm HCO}^+/{\rm H^{13}CO^+}$ intensity ratios observed towards Heiles Cloud 2 (which contains TMC-1) \citep{CernGue87} and could play a significant role in the radiative transfer of HCN. }

Another possible contributing factor to the presence of the anomalies involves considering the allowed downward pathways of excitation to the lower rotational state hyperfine energy levels. We examined all possible downward pathways from the J=7 level to the J=0 level in HCN. There are 36 different allowed routes via the different hyperfine transitions from this upper level, J=7, to the lower J=1 level. Remarkably, there is only one pathway out of the 36 that leads to emission of the leftmost lower hyper�ne transition in HCN, the J=1$\rightarrow$0, F=0$\rightarrow$1 line (which is involved most in the two effects described above). Three of the J=3-2 component hyper�ne lines also only have one allowed route to them.
When we considered the same situation for $\rm N_2H^+$, the number of pathways leading to line excitations as a result of transitions to each of the lower hyperfine levels, J$\rm F_{1}$F = 011 and 012, is 2-2.5 times that leading to excitation due to transitions to the J$\rm F_{1}$F = 010 level, thereby exhibiting a similar pattern to HCN above.  {\bf In contrast, a similar exercise for a non-anomalous hyperfine species, $\rm C^{17}O$,} shows an even number of routes down to the lower state hyper�ne energy levels. Therefore, it seems reasonable to suppose that small changes to the transition probabilities due to line overlap will readily disturb excitation of the HCN J=1$\rightarrow$0, F=0$\rightarrow$1 line in particular as well as to transitions down to the J$\rm F_{1}$F=010 hyperfine level in the case of $\rm N_2H^+$. The $\rm C^{17}$O line profile is much more robust because the lines are assembled from multiple downward hyperfine excitations in an homologous manner, that is, the hyperfine lines in the lower rotational state are excited equally. This may be due to a steady and consistent decrease of the available hyperfine energy levels as the hyperfine energy ladder is dismounted. Thus, if the number of pathways of excitation coming from upper hyperfine energy states and going to lower hyperfine energy states from a given hyperfine state (defined as the reference hyperfine energy state) does not diminish regularly as a result of a particular framework of hyperfine energy states, this will lead to a magnification of preferred routes of excitation down to the J=1 hyperfine levels (as is seen in HCN and $\rm N_2H^{+}$), perturbations of which result in hyperfine anomalies. It is important to note that this is a simplified argument that only addresses the downward excitation of available hyperfine energy states governed by the selection rules.  

It is worth noting that $\rm N_2H^+$ is also reported to exhibit anomalous hyperfine spectral components towards some sources \citep{Keto10,Daniel07} but not in G333. Since ${\rm N_2H^+}$ is found to be concentrate in the centre of dense quiescent cores and not participating in the large scale dynamical phenomenon traced in G333 by species such as HCN, this again points to a velocity doppler shifting causing line overlap being the likely cause of the anomalies.

\subsection{Observational implications}
Our study has shown that the hyperfine anomalies are common in star forming regions. If the line overlap hypothesis discussed above is correct, what are the implications for the use of HCN data in such regions? 

Clearly, measuring opacities by examining the relative intensities of the HCN hyperfine lines alone is unsafe and alternative methods should be used. HCN can still be used cautiously as a dynamical tracer species however. In regions where the turbulent velocity is low enough to resolve the HCN J=1$\rightarrow$0 hyperfine structure and discern self-absorption effects, then since the F=0$\rightarrow$1 appears to be most easily disturbed by line overlap effects, it will present a line profile shape that could be a complicated mixture of dynamical and selective absorption or emission effects. The line width may be affected too. Thus this line is an unreliable dynamical tracer in any source exhibiting resolved HCN (1-0) hyperfine anomalies and the central component should be used instead. In more turbulent sources such as in massive star formation, the J=1$\rightarrow$0, F=0$\rightarrow$1 hyperfine will give an inaccurate measurement of the line width (but of course a lower abundance species without hyperfine structure would be selected for such a measurement). In summary, it would seem prudent to simply exclude the HCN J=1$\rightarrow$0, F=0$\rightarrow$1 hyperfine line from any quantative calculation.

For HCN J=3$\rightarrow$2 for low mass, low turbulent width sources where the hyperfine lines are partially resolved, the central component should be reliable as a dynamical tracer since it is composed of three blended lines with many possible downward transitions to them. However, since the leftmost hyperfine is typically suppressed and the rightmost satellite boosted, then in circumstances where the lines are not quite resolved the velocity centroid may be miscalculated as lying between the central peak and rightmost satellite giving an error of order $\sim 1~{\rm km~s^{-1}}$ (half the separation of the central peak and rightmost satellite). In massive star forming regions, the possible presence of the anomalies mean that it should not be assumed that this line is composed of the central component alone. As discussed in section 4.4, the underlying hyperfine structure must be included to attempt to model infall and outflow using HCN J=3$\rightarrow$2 in massive star forming regions (Carolan et al 2009) and if the anomalies are present in the way found for low mass sources then the line shape will be distorted. In particular, constraining the degree of infall by measurements of the degree of line splitting (such as via the method of Mardones et al 1997; see Wu \& Evans 2003 for an example) is likely to result in an overestimate compared with using a species like ${\rm HCO^+}$. The magnitude of the difference in infall measurements between HCN and an alternative tracer species will be of order $\sim 2~{\rm km~s^{-1}}$, the separation of the central peak and rightmost satellite.

\subsection{Radiative transfer and future work}

A full detailed radiative transfer calculation of the line overlap hypothesis is clearly required to verify if the qualitative explanation above is correct. Line overlap, quite often termed line fluorescence has been successfully used at other wavelengths to explain, for example He~{\sc ii} optical line emission from ionized nebulae \citep{KastBhat90} and X-ray Fe lines very close to accreting massive black holes \citep{Fabetal00}. {\bf Line overlap is very difficult to calculate numerically and is typically not available in publically released radiative transfer codes. However, individual studies have succeeded in using line overlap to model the hyperfine line anomalies. Gonzalez-Alfonso \& Cernicharo (1993) \nocite{gonzalez93}used a modified Monte Carlo method to analyse the hyperfine anomalies in terms of the scattering approach described above. Daniel et al.\@~(2007) \nocite{Daniel07} have analysed the hyperfine component intensities of $\rm N_2H^+$ for several clouds by deriving a consistent set of hyperfine collisional coefficients for this molecule \citep{Daniel05} as well as carrying out nonlocal radiative transfer calculations of observational data using a large velocity gradient model \citep{Daniel06}. Recently, \citet{Daniel08} implemented the Gauss-Seidel algorithm in spherical geometry and included the case of line overlap of hyperfine transitions.

As part of our future work, we are modifying our RT code MOLLIE to use the \citep{Rybick92} line overlap algorithm in order to model the HCN anomalies. Our intention is to carry out a constraining of the parameter space giving rise to hyperfine anomalies in each of the two rotational transitions tabulated in Table \ref{tab:HCNcoreprops}}. It is hoped that the physical conditions common to our full contingent of twenty-eight starless cores will allow for such a study to be done at two individual transitions, while our database of 65 starless cores in the HCN J=1$\rightarrow$0 transition will allow for the study to be carried out extensively at the lower transition. Such a project will attempt to identify the physical characteristics of the star-forming core that gives rise to the anomalous intensities. This will focus on physical parameters such as the density, thermal as well as turbulent contributions to the line width, the velocity field of the core as well as the temperature gradient. Such constraining will involve the modelling of each core in multiple transitions of several different species with a 3D molecular radiative transport code \citep{Keto04, Redman02}.

\section{Conclusions}
While the existence of HCN anomalies have long been recognised in the J=1$\rightarrow$0 line \citep[e.g. for TMC\-1,][]{Walms82} for low mass star forming sources, our survey shows for the first time that the J=3$\rightarrow$2 can also be dramatically anomalous (see Figs. 1 and 2) and that these anomalies are common. Our study shows that massive star forming regions also exhibit hyperfine anomalies in the J=1$\rightarrow$0 line. These anomalies, in contrast to those of the low mass star forming sources, are apparent in line widths rather than line strengths. 

We favour a line overlap effect for the origin of the anomalies. It is likely that in higher rotational levels, where the hyperfine lines are more closely bunched, line overlap leads to preferred and suppressed radiative decay routes down to the lowest energy levels, where the hyperfines are widely separated, therefore, emphasising the disproportionate intensities. An attraction of the line overlap hypothesis is that is not strongly dependent on cloud column density since it is the form of the thermal, velocity and turbulent widths in the cloud that are the trigger for the anomalies. In a future paper we will carry out a full radiative transfer calculation of the hyperfine spectrum of HCN.

\section*{Acknowledgements}
We thank the referee for their report which led to an improved paper. We thank Jonathan Rawlings, David Williams, Eric Keto, Tigran Khanzadyan, Paul Jones, Michael Burton, Indra Bains and Jonathan Tennyson for useful discussions. Sincere gratitude is given to Dr. Jungjoo Sohn for making available her HCN J=1$\rightarrow$0 data for 65 starless cores and to Dr. Shuro Takano for providing HCN data for the TMC-1 starless core.

RML received financial support from the Irish Research Council for Science, Engineering and Technology (IRCSET). MPR acknowledges funding from a Science Foundation Ireland Research Frontiers Program grant (insert number). We thank the technical staff of the JCMT for obtaining the observations and especially Dr. Iain Coulson for assistance provided during the data collecting 
stage. The JCMT is operated by The Joint Astronomy Centre on behalf of the Science and Technology Facilities 
Council of the United Kingdom, the Netherlands Organisation for Scientific Research, and the National Research 
Council of Canada.

NL acknowledges partial support from Center of Excellence in Astrophysics and Associated Technologies (PFB 06) and Centro de Astrof\'{i}sica FONDAP\,15010003. NL's postdoctoral position at CEA/Irfu was funded by the Ile-de-France Region. The Mopra Telescope and ATCA are part of the Australia Telescope and are funded by the Commonwealth of Australia for operation as National Facility managed by CSIRO. The UNSW-MOPS Digital Filter Bank used for the observations with the Mopra Telescope was provided with support from the University of New South Wales, Monash University, University of Sydney and Australian Research Council.

BOD acknowledges STFC (UK). BOD was supported by funding from the Gates Cambridge Trust.

\begin{landscape}

\begin{table}
\centering
\begin{threeparttable}

\caption{Starless Cores Observed in HCN J=3$\rightarrow$2}

\begin{tabular}{l c c r @{} l c c c c c c c c c c c c}

\hline \hline
&&&&&&&&&&&\multicolumn{3}{c}{Detection Statistics}\\[-1ex]
\multicolumn{2}{c}{\raisebox{1.5ex}{\textsc{Source}}} & \raisebox{1.5ex}{R.A. (J2000.0)} & \multicolumn{2}{c}{\raisebox{1.5ex}{Decl. (J2000.0)}} & \raisebox{1.5ex}{Distance} & \raisebox{1.5ex}{Radius} & \raisebox{1.5ex}{$\rm V_{N_2H^+}$} & \raisebox{1.5ex}{$\tau_{\rm 225GHz}$} & \raisebox{1.5ex}{$\rm t_{int}$} & \raisebox{1.5ex}{TIF} & [D]/[ND] & $\sigma_{\rm T^{*}_{A}}$ & Noise Flux\\[-1ex]
&&&&& \raisebox{1.5ex}{(pc)} & \raisebox{1.5ex}{(pc)} & \raisebox{1.5ex}{(km$\rm s^{-1}$)} && \raisebox{1.5ex}{(hrs)} & \raisebox{1.5ex}{(Kkm$\rm s^{-1}$)} & & (K) & (Kkm$\rm s^{-1}$)\\
\multicolumn{2}{c}{\raisebox{1ex}{(1)}} & \raisebox{1ex}{(2)} & \multicolumn{2}{c}{\raisebox{1ex}{(3)}} & \raisebox{1ex}{(4)} & \raisebox{1ex}{(5)} & \raisebox{1ex}{(6)} & \raisebox{1ex}{(7)} & \raisebox{1ex}{(8)} & \raisebox{1ex}{(9)} & \raisebox{1ex}{(10)} & \raisebox{1ex}{(11)} & \raisebox{1ex}{(12)}\\[-0.5ex]
\toprule



L1498        & & 04 10 51.5 &  		    &25 09 58 & 140 & 0.057 & 7.82 & 0.193 & 0.67 & 0.313 & D & 0.089 & 0.038\\
L1495AN      & & 04 18 31.8 &  		    &28 27 30 & 140 & 0.076 & 7.30 & 0.089 & 0.83 & 0.280 & D & 0.041 & 0.017\\
L1521B       & & 04 24 12.7 &		    &26 36 53 & 140 & 0.044 & 6.42 & 0.255 & 2.00 & 0.374 & D & 0.042 & 0.113\\
B217-2	     & & 04 28 08.6 &		    &26 20 53 & 140 & 0.042 & 6.84 & 0.343 & 2.00 & 0.302 & D & 0.033 & 0.033\\
L1521F       & & 04 28 39.8 &  		    &26 51 35 & 140 & 0.045 & 6.49 & 0.089 & 0.66 & 0.295 & D & 0.065 & (0.181)\\
TMC-2        & & 04 32 49.0 &               &24 25 12 & 140 & 0.024 & 6.21 & 0.231 & 2.00 & 0.305 & D & 0.028 & 0.003 \\
CB22         & & 04 40 39.9 &  		    &29 52 59 & 140 & 0.036 & 5.97 & 0.204 & 2.00 & 0.098 & D & 0.029 & (0.057)\\
TMC-1        & & 04 41 33.0 &  		    &25 44 44 & 140 & 0.054 & 5.86 & 0.202 & 1.19 & 0.175 & D & 0.030 & 0.013\\
L1527B-1     & & 04 41 33.0 &               &25 46 24 & 140 & 0.045 & 5.90 & 0.216 & 2.00 & 0.376 & D & 0.065 & 0.006\\
CB23	     & & 04 43 31.5 &		    &29 29 11 & 140 & 0.034 & 6.04 & 0.238 & 2.00 & 0.117 & D & 0.032 & (0.0087)\\
L1507A       & & 04 42 38.6 &  		    &29 43 45 & 140 & 0.023 & 6.20 & 0.227 & 2.00 & 0.159 & D & 0.050 & (0.041)\\
L1517B       & & 04 55 18.8 &  		    &30 38 04 & 140 & 0.050 & 5.77 & 0.024 & 0.83 & 0.189 & D & 0.035 & 0.003\\
L1544        & & 05 04 14.9 &  		    &25 11 08 & 140 & 0.047 & 7.12 & 0.049 & 0.83 & 0.347 & D & 0.047 & 0.027\\
L1512        & & 05 05 09.7 &  		    &32 43 09 & 140 & 0.042 & 7.12 & 0.089 & 0.83 & 0.095 & D & 0.037 & (0.001)\\
L1552        & & 05 17 37.4 &  		    &26 05 28 & 140 & 0.039 & 7.64 & 0.204 & 2.67 & 0.369 & D & 0.029 & 0.037\\
L1582A       & & 05 32 03.4 &               &12 31 05 & 140 & 0.023 &10.21 & 0.246 & 2.00 & 0.079 & ND & 0.028 & (0.007)\\
L1622A2      & & 05 54 38.8 &      	    &01 53 44 & 500 & 0.035 & 1.10 & 0.049 & 0.83 & 0.878 & D & 0.042 & 0.064\\
L1622A1      & & 05 54 53.5 & \hspace{12pt} &01 57 24 & 500 & 0.044 & 1.15 & 0.237 & 2.67 & 0.343 & D & 0.034 & 0.005\\
L134A	     & & 15 53 33.1 &              -&04 35 26 & 165 & 0.019 & 2.68 & 0.182 & 2.00 & (0.024) & ND & 0.030 & 0.039\\
L1696A       & & 16 28 31.4 & 		   -&24 19 08 & 165 & 0.041 & 3.37 & 0.233 & 4.00 & 0.165 & D & 0.031 & 0.069\\
L1696B       & & 16 28 59.4 & 		   -&24 20 43 & 165 & 0.035 & 3.29 & 0.204 & 1.33 & 0.187 & D & 0.041 & 0.022\\
L1704-1      & & 16 30 50.7 & 		   -&23 42 14 & 165 & 0.032 & 2.67 & 0.262 & 2.00 & 0.040 & ND & 0.037 & (0.091)\\
L1689B       & & 16 34 45.9 & 		   -&24 37 51 & 165 & 0.029 & 3.49 & 0.198 & 2.00 & 0.130 & ND & 0.028 & 0.054\\
L204C-2      & & 16 47 46.3 & 		   -&12 23 19 & 165 & 0.037 & 4.24 & 0.258 & 2.00 & 0.096 & D & 0.027 & 0.003\\
L204F        & & 16 47 48.5 & 		   -&11 56 06 & 165 & 0.033 & 4.26 & 0.171 & 2.00 & (0.019) & ND & 0.027 & (0.055)\\
L234E-S      & & 16 48 08.7 & 		   -&10 57 25 & 165 & 0.038 & 3.08 & 0.186 & 2.00 & 0.242 & D & 0.050 & 0.085\\
L63          & & 16 50 15.6 & 		   -&18 05 16 & 165 & 0.028 & 5.78 & 0.277 & 2.00 & 0.181 & ND & 0.029 & (0.048)\\ 
L462-2       & & 18 07 36.3 & 		   -&04 40 29 & 200 & 0.041 & 7.90 & 0.294 & 1.33 & 0.252 & D & 0.026 & 0.052\\
L492         & & 18 15 46.1 & 		   -&03 46 13 & 200 & 0.038 & 7.72 & 0.195 & 4.00 & 0.524 & D & 0.031 & 0.007\\
L673-7       & & 19 21 35.6 &   	    &11 21 14 & 300 & 0.029 & 7.12 & 0.227 & 4.00 & 0.485 & D & 0.022 & 0.032\\
L694-2       & & 19 41 04.6 &  		    &10 57 02 & 250 & 0.032 & 9.57 & 0.210 & 3.33 & 0.296 & D & 0.022 & 0.014\\
L1148        & & 20 41 11.0 &  		    &67 20 35 & 325 & 0.027 & 2.60 & 0.285 & 2.00 & (0.122) & ND & 0.043 & 0.044\\
L1155C1      & & 20 43 30.1 &  		    &67 42 52 & 325 & 0.030 & 2.68 & 0.180 & 2.00 & 0.123 & D & 0.030 & 0.013\\
L944-2       & & 21 17 46.9 &  		    &43 18 20 & 700 & 0.046 & 5.35 & 0.191 & 2.67 & 0.096 & D & 0.023 & 0.019\\
L1197        & & 22 37 02.4 &  		    &58 57 21 & 400 & 0.026 & -3.16 & 0.328 & 3.33 & 0.316 & D & 0.027 & 0.006\\
CB246-2      & & 23 56 49.2 &  		    &58 34 29 & 140 & 0.030 & -0.84 & 0.223 & 2.00 & 0.104 & D & 0.031 & 0.026\\

\bottomrule

\end{tabular}

\begin{tablenotes}
\item[] \scriptsize{NOTES.-- (1) - observed starless cores in order of right ascension (R.A.); (2)-(3) - adopted coordinates toward peak centre of dust emission \citep[catalog of][in J2000.0]{Lee99}. (4) - distances towards observed sources; (5) - outermost contour radii of 350$\mu$m dust emission ($\ge$3$\sigma$ level) toward observed sources; (6) - the velocity of the optically thin $\rm N_{2}H^{+}$ main hyperfine (J,$\rm F_{1}$,F = 1,2,3$\rightarrow$0,1,2) component \citep{Sohn07}; (7) - median optical depth of atmosphere at 225GHz measured by the Caltech Submillimeter Observatory (CSO); effective integration time of the reduced spectrum for each of the cores in (8); (9) - representative flux (or Total Integrated Flux in Kk$\rm m^{-1}$) for HCN spectra in region of width $V_{\rm N_2H^{+}}\pm10\rm km\rm s^{-1}$. (10) specifies a Detection [D] or a Non-Detection [ND] with respect to the corresponding source. (11) - standard deviation of noise brightness in region devoid of emission, 20km$\rm s^{-1}$ in width, comparable with (6) of Table 2 and gives some idea of the degree of noise apparent in weather band 5 observations at the JCMT. In order to allow a computation of the signal-to-noise ratio (S/N) for each of the spectra, the noise flux in the same region assessed in column eleven is listed in column 12 (in K km$\rm s^{-1}$).
The units of right ascension (R.A.) are in hours, minutes, and seconds, and units of declination (Decl.) are in degrees, arcminutes and arcseconds.
Values in braces in (9) and (12) represent negative values found in analysis of corresponding spectra.
$\rm^{a}D$istances are attributed to the following references: \citep{LeeMyerTaf01} - L1498, L1495A-N, L1521F, TMC-1, L1507A, L1517B, L1512, L1544, L134A, L1689B, L234E-S, L492, L694-2 $\&$ CB246-2; \citep{Chini81} (Ophiuchus) - L1696A, L1696B, L204C-2, L204F, L1704-1, L63; \citep{Maddetal86} - L1622A-1, L1622A-2; \citep{Dameetal85} (Aquila Rift [200pc], Cyg Rift [700pc], Vul Rift [400pc]) - L462-2, L1197, L944-2; \citep{Straizys92} - L1148, L1155C-1; \citep{Fellietal92} - L673-7; \citep{Elias78} - L1552, CB22. 
$\rm^{b}V$alues for the opacity reflect the observing conditions under which the observations were taken [refer to text]. Values in parentheses represent negative quantities.}

\end{tablenotes}


\label{tab:CoreLineParams}
\end{threeparttable}
\end{table}
\end{landscape}

\begin{landscape}
\begin{table}
\centering
\begin{threeparttable}
\caption{The Line Parameters of the Starless Cores Observed in HCN}\label{tab:HCNcoreprops}

\begin{tabular}{lc|cccc||cccc}


\hline \hline
&&& \multicolumn{3}{c||}{$\int T^{*}_A \Delta v$, J=1$\rightarrow$0 (Kkm$\rm s^{-1}$)} && \multicolumn{3}{c}{$\int T^{*}_A \Delta v$, J=3$\rightarrow$2 (Kkm$\rm s^{-1}$)} \\[-1ex]
&& \raisebox{1.5ex}{$\rm T_{A}^{*}$(J,F=1,2$\rightarrow$0,1)} &\multicolumn{3}{c||}{\raisebox{1.0ex}{\bf{------------------------------------}}}& \raisebox{1.5ex}{$\rm T_{A}^{*}$(J,F=3,4$\rightarrow$2,3)}& \multicolumn{3}{c}{\raisebox{1.0ex}{\bf{------------------------------------}}}\\[-2.5ex]
\multicolumn{2}{c|}{\raisebox{1.5ex}{\textsc{Source}}} & (K) & F=0$\rightarrow$1 & F=2$\rightarrow$1 & F=1$\rightarrow$1 & (K) & $\rm\Delta F=0^{-}$ & $\rm\Delta F=1$ & $\rm\Delta F=0^{+}$\\ 
\multicolumn{2}{c|}{(1)} & (2) & (3) & (4) & (5) & (6) & (7) & (8) & (9) \\
\toprule

L1498        &         & 1.00 & 0.171$\pm$0.014 & 0.175$\pm$0.014 & 0.147$\pm$0.012 & 0.35 & 0.017$\pm$0.000 & 0.140$\pm$0.003 & 0.101$\pm$0.002 \\
L1495AN      &         & 0.65 & 0.218$\pm$0.017 & 0.305$\pm$0.024 & 0.184$\pm$0.015 & 0.18 & 0.020$\pm$0.000 & 0.165$\pm$0.003 & 0.137$\pm$0.003 \\
L1521B       &	       & 0.36 & 0.185$\pm$0.015 & 0.224$\pm$0.018 & 0.196$\pm$0.016 & 0.17 & 0.015$\pm$0.000 & 0.084$\pm$0.002 & 0.092$\pm$0.002 \\
B217-2       &         & 1.09 & 0.336$\pm$0.027 & 0.495$\pm$0.040 & 0.306$\pm$0.024 & 0.23 & 0.046$\pm$0.001 & 0.179$\pm$0.004 & 0.074$\pm$0.001 \\
L1521F       &         & 0.73 & 0.272$\pm$0.022 & 0.296$\pm$0.024 & 0.237$\pm$0.019 & 0.71 & 0.046$\pm$0.001 & 0.253$\pm$0.005 & 0.155$\pm$0.003 \\
TMC-2        &         & 1.50 & 0.234$\pm$0.019 & 0.518$\pm$0.041 & 0.332$\pm$0.027 & 0.25 & 0.053$\pm$0.001 & 0.172$\pm$0.003 & 0.118$\pm$0.002 \\
CB22         &         & 0.77 & 0.150$\pm$0.012 & 0.259$\pm$0.021 & 0.157$\pm$0.013 & 0.13 & 0.012$\pm$0.000 & 0.072$\pm$0.001 & 0.013$\pm$0.000 \\
TMC-1        &         & 1.08 & 0.390$\pm$0.031 & 0.355$\pm$0.028 & 0.304$\pm$0.024 & 0.20 & 0.023$\pm$0.000 & 0.104$\pm$0.002 & 0.075$\pm$0.002 \\
L1527B-1     &         & 0.57 & 0.260$\pm$0.021 & 0.240$\pm$0.019 & 0.180$\pm$0.014 & 0.26 & 0.066$\pm$0.001 & 0.185$\pm$0.004 & 0.111$\pm$0.002 \\
CB23         &         & 0.78 & 0.122$\pm$0.010 & 0.198$\pm$0.016 & 0.124$\pm$0.010 & 0.09 & 0.002$\pm$0.000 & 0.049$\pm$0.001 & 0.068$\pm$0.001 \\
L1507A       &         & 0.74 & 0.190$\pm$0.015 & 0.224$\pm$0.018 & 0.139$\pm$0.011 & 0.12 & 0.039$\pm$0.001 & 0.127$\pm$0.002 & 0.041$\pm$0.001 \\
L1517B       &         & 0.67 & 0.167$\pm$0.013 & 0.237$\pm$0.019 & 0.124$\pm$0.010 & 0.24 & 0.009$\pm$0.000 & 0.090$\pm$0.002 & 0.053$\pm$0.001 \\
L1544        &         & 1.27 & 0.305$\pm$0.024 & 0.258$\pm$0.021 & 0.275$\pm$0.022 & 0.57 & 0.063$\pm$0.001 & 0.225$\pm$0.004 & 0.176$\pm$0.003 \\
L151$\rm 2^{\ddagger}$        &         & 0.73 & 0.128$\pm$0.010 & 0.202$\pm$0.016 & 0.125$\pm$0.010 & 0.23 & 0.005$\pm$0.000 & 0.083$\pm$0.002 & 0.055$\pm$0.001 \\
L1552        &         & 0.69 & 0.139$\pm$0.011 & 0.137$\pm$0.011 & 0.117$\pm$0.009 & 0.27 & 0.052$\pm$0.001 & 0.085$\pm$0.002 & 0.089$\pm$0.002 \\
L1622A2      &         & 0.76 & 0.336$\pm$0.027 & 0.496$\pm$0.040 & 0.243$\pm$0.019 & 0.57 & 0.251$\pm$0.005 & 0.382$\pm$0.008 & 0.092$\pm$0.002 \\
L1622A1      &         & 0.90 & 0.322$\pm$0.026 & 0.406$\pm$0.032 & 0.252$\pm$0.020 & 0.27 & 0.072$\pm$0.001 & 0.216$\pm$0.004 & 0.104$\pm$0.002 \\
L1696A       &         & 0.89 & 0.310$\pm$0.025 & 0.444$\pm$0.036 & 0.302$\pm$0.024 & 0.33 & 0.031$\pm$0.001 & 0.175$\pm$0.003 & 0.088$\pm$0.002 \\
L1696B       &         & 0.41 & 0.242$\pm$0.019 & 0.322$\pm$0.026 & 0.211$\pm$0.017 & 0.23 & 0.042$\pm$0.001 & 0.154$\pm$0.003 & 0.104$\pm$0.002 \\
L204C-$\rm 2^{\dagger}$      &         & 0.77 & 0.251$\pm$0.020 & 0.604$\pm$0.048 & 0.332$\pm$0.027 & 0.17 & 0.001$\pm$0.000 & 0.155$\pm$0.003 & 0.005$\pm$0.000 \\
L234E-$\rm S^{\dagger}$      &         & 0.62 & 0.183$\pm$0.015 & 0.343$\pm$0.027 & 0.248$\pm$0.020 & 0.16 & 0.005$\pm$0.000 & 0.092$\pm$0.002 & 0.042$\pm$0.001 \\
L462-$\rm 2^{\dagger}$       &         & 0.42 & 0.185$\pm$0.015 & 0.256$\pm$0.020 & 0.307$\pm$0.024 & 0.13 & 0.019$\pm$0.000 & 0.103$\pm$0.002 & 0.058$\pm$0.001 \\
L492         &         & 0.69 & 0.279$\pm$0.022 & 0.384$\pm$0.031 & 0.286$\pm$0.023 & 0.46 & 0.066$\pm$0.001 & 0.225$\pm$0.004 & 0.143$\pm$0.003 \\
L673-7       &         & 0.63 & 0.200$\pm$0.016 & 0.490$\pm$0.039 & 0.249$\pm$0.020 & 0.41 & 0.035$\pm$0.001 & 0.330$\pm$0.007 & 0.144$\pm$0.003 \\
L694-2       &         & 0.17 & 0.108$\pm$0.009 & 0.095$\pm$0.010 & 0.107$\pm$0.009 & 0.22 & 0.045$\pm$0.001 & 0.154$\pm$0.003 & 0.051$\pm$0.001 \\
L1155C1      &         & 0.51 & 0.135$\pm$0.011 & 0.182$\pm$0.015 & 0.131$\pm$0.010 & 0.14 & 0.006$\pm$0.000 & 0.073$\pm$0.001 & 0.048$\pm$0.001 \\
L944-2       &         & 0.37 & 0.076$\pm$0.006 & 0.165$\pm$0.013 & 0.089$\pm$0.007 & 0.17 & 0.022$\pm$0.000 & 0.082$\pm$0.002 & 0.033$\pm$0.001 \\
L1197        &         & 0.26 & 0.125$\pm$0.010 & 0.132$\pm$0.011 & 0.105$\pm$0.008 & 0.36 & 0.096$\pm$0.002 & 0.149$\pm$0.003 & 0.086$\pm$0.002 \\
CB246-$\rm 2^{\dagger}$      &	       & 0.42 & 0.123$\pm$0.010 & 0.261$\pm$0.021 & 0.163$\pm$0.013 & 0.12 & 0.021$\pm$0.000 & 0.062$\pm$0.001 & 0.019$\pm$0.000 \\

\bottomrule

\end{tabular}

\begin{tablenotes}
\item[] \scriptsize{NOTES.-- Columns (2) \& (6) detail the main hyperfine component brightness in each of the two rotational transitions in degrees Kelvin, respectively. Those values listed in column (2) are directly adapted from Table 1 of \citep{Sohn07}. Columns (3),(4) \& (5) reproduce the computed integrated emission for each hyperfine component of the J=1$\rightarrow$0 transition, that is, J,F = 1,0$\rightarrow$0,1, J,F = 1,2$\rightarrow$0,1, J,F = 1,1$\rightarrow$0,1 respectively. The same assessment was carried out for the higher rotational transition in columns (7), (8) \& (9) where the integrated emissions for the hyperfine branches $\Delta F = 0^{-}$ (leftward shifted (in velocity) hyperfine branch, the J,F=3,3$\rightarrow$2,3 component), $\Delta F = 1$ (central group of 3 hyperfine components, J,F=3,2$\rightarrow$2,1, 3,3$\rightarrow$2,2 \& 3,4$\rightarrow$2,3) \& $\Delta F$ = $0^{+}$ (rightward shifted (in velocity) hyperfine branch, the J,F=3,2$\rightarrow$2,2 component) are displayed, respectively.} 
\item[] \scriptsize{$\rm ^{\dagger}T$he leftmost component, J,F = 3,3$\rightarrow$2,3 for these sources was quite diminished and so a region of width equivalent to that of the rightmost component and with the correct offset was integrated.\\
$\rm ^{\ddagger}N$oisy spectrum shifted upward by an amount equivalent to presumed baseline in order to give correct integrated flux for leftmost component, J,F = 3,3$\rightarrow$2,3.}

\end{tablenotes}

\end{threeparttable}
\end{table}
\end{landscape}

\appendix

\section{Converged HCN J=1$\rightarrow$0 Hyperfine Component Analysis}

\begin{figure*}
\begin{center}
\includegraphics[width=450pt]{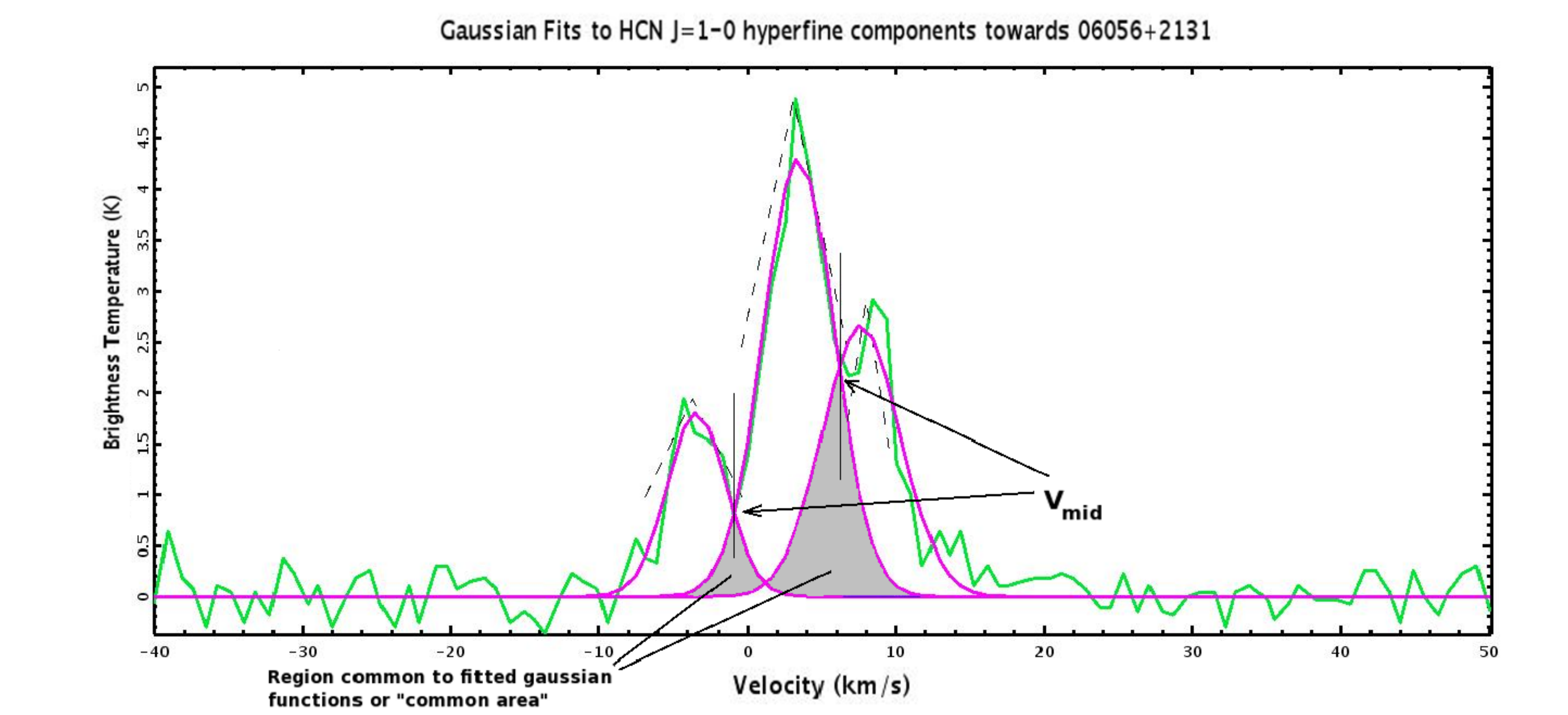}
\caption{Resultant gaussian fit to the J=1$\rightarrow$0 hyperfine components of the bright infrared source 06056+2131 \citep[source data from][]{Pirog99}.}
\label{fig:gaussianfit}
\end{center}
\end{figure*}

The technique that was employed to deal with the partially overlapped hyperfine components for the J=1$\rightarrow$0 transition is outlined as 
follows. Initially, for each component in an overlapped hyperfine spectrum, a gaussian approximation was implemented (using Starlink Spectral 
Analysis Tool, $\rm\textsc{splat}$) that took the following form:
\begin{equation}
\rm y(x) = A\exp{\left(-0.5\left[\frac{v-c}{\sigma}\right]^2\right)}
\label{eq:gaussapprox}
\end{equation}
where A is the scale height (peak height) of the approximated region (the gaussian peaks in  Fig.~\ref{fig:gaussianfit}), c is the central 
velocity-position of the gaussian-peak and $\rm\sigma$ is the gaussian width. Thus once the fitted gaussian function is satisfactory, there 
will be a unique set of values for A, c and $\rm\sigma$, based on the form of the original spectral component.

In our proposed exact calculation of the overlapped region, the area is equivalent to the sum of two contributions. Firstly, in 
Fig.~\ref{fig:gaussianfit}, the section of the common area  to the right of the line marked by $\rm v_{mid}$ ($\rm v_{mid}$ marks 
the velocity-position where the intensity of the overlapped area reaches a maximum) is equal to the integrated area over the complete 
gaussian on the LHS minus the integral over this gaussian function, $\rm y_1$(v), from the limits -$\rm\infty$ to $\rm v_{mid}$. Secondly, 
it is reasonably straightforward to deduce that the remaining section of the ``common area'' to the left of the line marked by $\rm v_{mid}$ 
is equal to the integrated area over the complete gaussian on the RHS minus the integral over this gaussian function, $\rm y_2$(v), from the 
limits $\rm v_{mid}$ to +$\rm\infty$. The overall area, $\rm\Sigma$, is then the sum of these two segments giving,

\begin{eqnarray}
\rm \Sigma &=& \left(\int_{-\infty}^{\infty}y_1(v)dv - \int_{-\infty}^{\rm v_{mid}} y_1(v)dv\right)\nonumber\\
&+& \left(\int_{-\infty}^{\infty}y_2(v)dv - \int_{\rm v_{mid}}^{\infty} y_2(v)dv\right).
\label{eq:intarea}
\end{eqnarray}

The integration over a gaussian function above or below a certain threshold $\rm v_{mid}$ is not available analytically, but the solution has 
a simple relationship to the error function, $\rm\mbox{Erf}$(v), or its complement, $\rm\mbox{Erfc}$(v) \citep[tabulated in mathematical 
tables,][]{Beyer91},

\begin{subequations}
\begin{eqnarray}
\rm\mbox{Erf}(v) &=& \frac{2}{\sqrt{\pi}}\int^v_0e^{-u^2}du\\
\rm\mbox{Erfc}(v) &=& \rm 1 - \mbox{Erf}(v)\nonumber\\
&=& \frac{2}{\sqrt{\pi}}\int_v^{\infty}e^{-u^2}du
\end{eqnarray}
\end{subequations}

It is the error function itself that plays an important role in each of the above integrals in Eq.~\eqref{eq:intarea}. Additionally, 
$\rm\mbox{Erf}$(0)=0 and $\rm\mbox{Erf}(\infty)$=1 aided in the calculations. Along with the unique set of constants that describe the 
form of the fitted gaussian function for each of the hyperfine spectral components, the quantity, $\rm v_{mid}$, crucial for the integrated 
emission calculation via Eq.~\eqref{eq:intarea}, had to be determined analytically for each set of overlapping gaussian components. In 
accordance with the above description, each of the gaussian approximated functions are represented by $\rm y_1$(v) and $\rm y_2$(v), 
respectively. Then, via Eq.~\eqref{eq:gaussapprox}, the calculation proceeded as follows:

\begin{equation}
\rm y_1(v) = \rm y_2(v)
\label{eq:overlapreg2}
\end{equation}
  
After much rearranging of Eq.~\eqref{eq:overlapreg2}, the following quadratic in v is achieved,

\begin{equation}
\begin{split}
\rm (\sigma_2^2-\sigma_1^2)v^2 + 2(\sigma_1^2c_2-\sigma_2^2c_1)v\\
-\left[2\sigma_1^2\sigma_2^2\ln{\left(\frac{A_1}{A_2}\right)}+\sigma_1^2c_2^2-\sigma_2^2c_1^2\right]=0
\label{eq:quadinv}
\end{split}
\end{equation}

Using the standard techniques for solving Eq.~\eqref{eq:quadinv}, two values are achieved, one of which lies between the values $\rm c_1$ 
and $\rm c_2$ each for the two gaussians respectively. This is $\rm v_{mid}$ in Eq.~\eqref{eq:intarea} above. The other value is another 
velocity-position at which the two overlapping gaussians intercept, but far from the vicinity of ``common area'' in Fig.~\ref{fig:gaussianfit} 
given above (this depends on the width of each of the respective gaussian functions). Neglecting the latter, the equation for $\rm\Sigma$ 
above can be solved routinely to give the area of the overlapped region. This area is apportioned to each of the overlapped spectral components 
(these spectral components are actually hyperfine branches) with respect to the ratio of the non-overlapping sections of these profiles, i.e. 
for the LHS component in Fig.~\ref{fig:gaussianfit} above, the non-overlapping area is given by $\rm\int_{-\infty}^{v_{mid}}y_1$(v)dv. The 
middle hyperfine branch in the HCN J=3$\rightarrow$2 transition overlaps both the left and right branches, and so the non-overlapping segment 
of this branch was determined by subtracting sections from each of the overlapped areas. Once the respective portions from each of the 
overlapped regions were distributed accordingly to each of the hyperfine components, the relative intensity ratios, $\rm R_{02}$ and $\rm R_{12}$ 
could be calculated. Fig.~\ref{fig:highmassrat} includes these ratios, each value subject to the analytical treatment described here.

\label{lastpage}


\begin{thebibliography}{}

\if{unused_refs}

\bibitem[\protect\citeauthoryear{Bentley et al.}{1993}]{Bent93}
    Bentley J.~A., Huang C.~M., Wyatt R. E., 1993, J. Chem. Phys., 98, 5207 

\bibitem[\protect\citeauthoryear{Beyer}{1991}]{1991Beyer} Beyer 
W.~H., 1991, csmt.book, 29

\bibitem[\protect\citeauthoryear{Boger \& Sternberg}{2005}]{BogStern05}
    Boger G. I., Sternberg A., 2005, ApJ, 632, 302

\bibitem[\protect\citeauthoryear{Chini}{1981}]{Chini81} 
    Chini, R., 1981, A\&A, 99, 346	

     

\bibitem[\protect\citeauthoryear{Suzuki et al.}{1992}]{Suzuki92}
    Suzuki, H., Yamamoto, S., Ohishi, M., Kaifu, N., Ishikawa, S., Hirahara,
    Y., Takano, S. 1992, ApJ, 392, 551 
  
\fi

\bibitem[\protect\citeauthoryear{Ahrens et al.}{2002}]{Ahrens02}
    Ahrens V., Lewen F., Takano S., Winnewisser S., Urban S., Negirev A. A., Koroliev A. N. 
    2002, Z. Naturforsch., 57\textbf{a}, 669

\bibitem[\protect\citeauthoryear{Alves, Lada, 
\& Lada}{2001}]{alves01} Alves J.~F., Lada C.~J., Lada E.~A., 2001, Natur, 409, 159

\bibitem[\protect\citeauthoryear{Bains et al.}{2006}]{bains06} 
Bains I., et al., 2006, MNRAS, 367, 1609 

\bibitem[\protect\citeauthoryear{Bensch et al.}{2001}]{Bensch01}
    Bensch F., Panis J.-F., Stutzki J., Heithausen A., Falgarone E., 2001, A\&A, 365, 275 

\bibitem[\protect\citeauthoryear{Beyer}{1991}]{Beyer91} Beyer 
W.~H., 1991, CRC standard mathematical tables and formulae

\bibitem[{{Boonman} {et~al.}(2001){Boonman}, {Stark}, {van der Tak}, {van
  Dishoeck}, {van der Wal}, {Sch{\"a}fer}, {de Lange}, \&
  {Laauwen}}]{boonman01}
{Boonman}, A.~M.~S., {Stark}, R., {van der Tak}, F.~F.~S., {et~al.} 2001, ApJL,
  553, L63

\bibitem[\protect\citeauthoryear{Cao et al.}{1993}]{1993Cao}
    Cao Y.~X., Zeng Q., Deguchi S., Kameya O., Kaifu N., 1993, AJ, 105, 1027
    
    \bibitem[\protect\citeauthoryear{Carolan et 
al.}{2008}]{carolan08} Carolan P.~B., Redman M.~P., Keto E., 
Rawlings J.~M.~C., 2008, MNRAS, 383, 705 

\bibitem[\protect\citeauthoryear{Carolan et 
al.}{2009}]{carolan09} Carolan P.~B., et al., 2009, MNRAS, 400, 
78 

\bibitem[\protect\citeauthoryear{Cernicharo et al.}{1984}]{Cerni84}
    Cernicharo J., Castets A., Duvert G., Guilloteau S., 1984, A\&A, 139, L13

\bibitem[\protect\citeauthoryear{Cernicharo \& Guelin}{1987}]{CernGue87}
    Cernicharo J., Guelin M., 1987, A\&A, 183, 10

\bibitem[\protect\citeauthoryear{Chini}{1981}]{Chini81} 
    Chini, R., 1981, A\&A, 99, 346	

\bibitem[{{Choi} {et~al.}(1999){Choi}, {Panis}, \& {Evans}}]{choi99}
{Choi}, M., {Panis}, J., \& {Evans}, II, N.~J. 1999, ApJS, 122, 519
 
 \bibitem[{{Crapsi} {et~al.}(2005){Crapsi}, {Caselli}, {Walmsley}, {Myers},
  {Tafalla}, {Lee}, \& {Bourke}}]{crapsi05}
{Crapsi}, A., {Caselli}, P., {Walmsley}, C.~M., {et~al.} 2005, ApJ, 619, 379
 
 \bibitem[\protect\citeauthoryear{Dame \& Thaddeus}
    {1985}]{Dameetal85} Dame T. M., Thaddeus P., 1985, ApJ, 297, 751 
    
\bibitem[\protect\citeauthoryear{Daniel \& Cernicharo}{2008}]{Daniel08}
    Daniel F., Cernicharo J., 2008, A\&A, 488, 1237

\bibitem[\protect\citeauthoryear{Daniel et al.}{2005}]{Daniel05}
    Daniel F., Dubernet M. L., Meuwly M., Cernicharo J., Pagani L., 2005, MNRAS, 363, 1083

\bibitem[\protect\citeauthoryear{Daniel et al.}{2006}]{Daniel06}
    Daniel F., Cernicharo J., Dubernet M. L., 2006, ApJ, 648, 461
   
\bibitem[\protect\citeauthoryear{Daniel et al.}{2007}]{Daniel07}
Daniel, F., Cernicharo, J., Roueff, E., Gerin, M., Dubernet, M. L.,  2007, ApJ, 667, 980

\bibitem[\protect\citeauthoryear{Elias}{1978}]{Elias78}
    Elias J. H., 1978, ApJ, 224, 453

\bibitem[\protect\citeauthoryear{Fabian et al.}{2000}]{Fabetal00}
    Fabian, A. C., Saunders J. S., Ettori S., Taylor G. B., Allen S. W.,
    Crawford C. S., Iwasawa K., Johnstone R. M., Ogle P. M., 2000, MNRAS, 318, L65 

\bibitem[\protect\citeauthoryear{Faure et al.}{2007}]{faure.et.al07} 
Faure A., Varambhia H.~N., Stoecklin T., Tennyson J., 2007, MNRAS, 382, 840 

\bibitem[\protect\citeauthoryear{Felli et al.}{1992}]{Fellietal92}
    Felli M., Palagi F., Tofani G., 1992, A\&A, 255, 293

\bibitem[\protect\citeauthoryear{Freed 
\& Mangum}{2005}]{Freed05} Freed K.~M., Mangum J.~G., 2005, AAS, 37, \#184.08

\bibitem[{{Friedel} {et~al.}(2005){Friedel}, {Remijan}, {Snyder}, {A'Hearn},
  {Blake}, {de Pater}, {Dickel}, {Forster}, {Hogerheijde}, {Kraybill},
  {Looney}, {Palmer}, \& {Wright}}]{friedel05}
{Friedel}, D.~N., {Remijan}, A.~J., {Snyder}, L.~E., {et~al.} 2005, ApJ, 630,
  623

\bibitem[{{Gao} \& {Solomon}(2004)}]{gao04}
{Gao}, Y. \& {Solomon}, P.~M. 2004, ApJS, 152, 63
   \bibitem[\protect\citeauthoryear{Goldsmith \& Irvine}{1986}]{Gold86}
    Goldsmith P. F., Irvine W. M., 1986, ApJ, 310, 383

\bibitem[\protect\citeauthoryear{Gottlieb et al.}{1975}]{Gott75}
    Gottlieb C. A., Lada C. J., Gottlieb E. W., Lilley A. E., Litvak M. M., 1975, ApJ, 202, 655

\bibitem[\protect\citeauthoryear{Gonzalez-Alfonso 
\& Cernicharo}{1993}]{gonzalez93} Gonzalez-Alfonso E., Cernicharo J., 1993, A\&A, 279, 506 

\bibitem[\protect\citeauthoryear{Guilloteau \& Baudry}{1981}]{Guill81}
    Guilloteau S., Baudry, A., 1981, A\&A, 97, 213 

\bibitem[{{Hennemann} {et~al.}(2009){Hennemann}, {Birkmann}, {Krause}, {Lemke},
  {Pavlyuchenkov}, {More}, \& {Henning}}]{hennemann09}
{Hennemann}, M., {Birkmann}, S.~M., {Krause}, O., {et~al.} 2009, ApJ, 693, 1379

\bibitem[{{Hirota} {et~al.}(1999){Hirota}, {Yamamoto}, {Kawaguchi}, {Sakamoto},
  \& {Ukita}}]{hirota99}
{Hirota}, T., {Yamamoto}, S., {Kawaguchi}, K., {Sakamoto}, A., \& {Ukita}, N.
  1999, ApJ, 520, 895

\bibitem[{{Hogerheijde} {et~al.}(2009){Hogerheijde}, {Qi}, {de Pater}, {Blake},
  {Friedel}, {Forster}, {Palmer}, {Remijan}, {Snyder}, \&
  {Wright}}]{hogerheijde09}
{Hogerheijde}, M.~R., {Qi}, C., {de Pater}, I., {et~al.} 2009, AJ, 137, 4837

\bibitem[{{Hunt} {et~al.}(1999){Hunt}, {Whiteoak}, {Cragg}, {White}, \&
  {Jones}}]{hunt99}
{Hunt}, M.~R., {Whiteoak}, J.~B., {Cragg}, D.~M., {White}, G.~L., \& {Jones},
  P.~A. 1999, MNRAS, 302, 1

\bibitem[\protect\citeauthoryear{Irvine \& Schloerb}{1984}]{Irvine84}
    Irvine W. M., Schloerb F. P., 1984, ApJ, 282, 516

\bibitem[\protect\citeauthoryear{Kastner \& Bhatia}{1990}]{KastBhat90}
    Kastner S. O., Bhatia A. K., 1990, ApJ, 362, 745

\bibitem[\protect\citeauthoryear{Keto et al.}{2004}]{Keto04}
    Keto E., Rybicki G. B., Bergin E. A., Plume R., 2004, ApJ, 613, 355

\bibitem[\protect\citeauthoryear{Keto \& Rybicki}{2010}]{Keto10}
    Keto E., Rybicki G. B., 2010, ApJ, 716, 1315

\bibitem[{{Kohno} {et~al.}(2003){Kohno}, {Ishizuki}, {Matsushita},
  {Vila-Vilar{\'o}}, \& {Kawabe}}]{kohno03}
{Kohno}, K., {Ishizuki}, S., {Matsushita}, S., {Vila-Vilar{\'o}}, B., \&
  {Kawabe}, R. 2003, PASJ, 55, L1
    
\bibitem[\protect\citeauthoryear{Kwan \& Scoville}{1975}]{Kwan75}
    Kwan J., Scoville N., 1975, ApJL, 195, L85

\bibitem[\protect\citeauthoryear{Ladd et al.}{2005}]{Ladd05}
    Ladd N., Purcell C., Wong T., Robertson S., 2005, PASA, 22, 65

\bibitem[\protect\citeauthoryear{Lapinov}{1989}]{Lapinov89}
    Lapinov, A.~V., 1989, SvA, 33, 132
    
\bibitem[\protect\citeauthoryear{Lee et al.}{1999}]{Lee99}
    Lee C. W., Myers P. C., Tafalla M., 1999, ApJ, 526, 788

\bibitem[\protect\citeauthoryear{Lee et al.}{2001}]{LeeMyerTaf01}
    Lee C. W., Myers P. C., Tafalla M., 2001, ApJS, 136, 703

\bibitem[\protect\citeauthoryear{Lo et al.}{2009}]{Lo09}
    Lo N., Cunningham M.~R., Jones P.~A., Bains I., Burton M.~G., 
    Wong T., Muller E., Kramer C., Ossenkopf V., Henkel C., Deragopian G., 
    Donnelly S., Ladd E.~F., 2009, MNRAS, 395, 1021

\bibitem[\protect\citeauthoryear{Maddalena et al.}{1986}]{Maddetal86}
    Maddalena R. J., Morris M., Moscowitz J., Thaddeus P., 1986, ApJ, 303, 375

\bibitem[\protect\citeauthoryear{Mardones et 
al.}{1997}]{mardones97} Mardones D., Myers P.~C., Tafalla M., 
Wilner D.~J., Bachiller R., Garay G., 1997, ApJ, 489, 719 

\bibitem[{{Marten} {et~al.}(2002){Marten}, {Hidayat}, {Biraud}, \&
  {Moreno}}]{marten02}
{Marten}, A., {Hidayat}, T., {Biraud}, Y., \& {Moreno}, R. 2002, Icarus, 158,
  532

\bibitem[\protect\citeauthoryear{Mookerjea et 
al.}{2004}]{mookerjea04} Mookerjea B., Kramer C., Nielbock M., Nyman L.-{\AA}., 2004, A\&A, 426, 119

\bibitem[\protect\citeauthoryear{Papadopoulos}{2007}]{Papa07}
    Papadopoulos, P. P., 2007, ApJ, 656, 792

\bibitem[{{Park} {et~al.}(1999){Park}, {Kim}, \& {Minh}}]{park99}
{Park}, Y., {Kim}, J., \& {Minh}, Y.~C. 1999, ApJ, 520, 223

\bibitem[\protect\citeauthoryear{Pirogov}{1999}]{Pirog99}
    Pirogov L., 1999, A\&A, 348, 600

\bibitem[\protect\citeauthoryear{Redman et al.}{2002}]{Redman02}
    Redman M. P., Rawlings J. M. C., Nutter D. J., Ward-Thompson D., Williams D. A., 2002, MNRAS, 337, L17 

\bibitem[\protect\citeauthoryear{Redman et al.}{2004}]{Redman04} 
Redman M.~P., Keto E., Rawlings J.~M.~C., Williams D.~A., 2004, MNRAS, 352, 
1365 

\bibitem[\protect\citeauthoryear{Redman, Keto, 
\& Rawlings}{2006}]{Redman06} Redman M.~P., Keto E., Rawlings J.~M.~C., 2006, MNRAS, 370, L1 

\bibitem[\protect\citeauthoryear{Rybicki \& Hummer}{1992}]{Rybick92}
    Rybicki G. B., Hummer D. G., 1992, A\&A, 262, 209

\bibitem[{{Sarrasin} {et~al.}(2010){Sarrasin}, {Abdallah}, {Wernli}, {Faure},
  {Cernicharo}, \& {Lique}}]{sarrasin10}
{Sarrasin}, E., {Abdallah}, D.~B., {Wernli}, M., {et~al.} 2010, MNRAS, 404, 518

\bibitem[{{Schilke} {et~al.}(2000){Schilke}, {Mehringer}, \&
  {Menten}}]{schilke00}
{Schilke}, P., {Mehringer}, D.~M., \& {Menten}, K.~M. 2000, ApJL, 528, L37

\bibitem[{{Schilke} \& {Menten}(2003)}]{schilke03}
{Schilke}, P. \& {Menten}, K.~M. 2003, ApJ, 583, 446

\bibitem[\protect\citeauthoryear{Snyder \& Buhl}{1971}]{SnyBuhl71}
    Snyder L. E., Buhl D., 1971, ApJ, 163, L47

\bibitem[\protect\citeauthoryear{Sohn et al.}{2007}]{Sohn07}
    Sohn J., Lee C. W., Park Y. S., Lee H. M., Myers P. C., Lee Y., 2007, ApJ, 664, 928

\bibitem[\protect\citeauthoryear{Straizys et al.}{1992}]{Straizys92}
    Straizys V., Cernis K., Kazlauskas A., Meistas E., 1992, BaltA, 1, 149

\bibitem[\protect\citeauthoryear{Stahler 
\& Yen}{2010}]{stahler10} Stahler S.~W., Yen J.~J., 2010, MNRAS, 407, 2434 

\bibitem[\protect\citeauthoryear{Tafalla et al.}{2006}]{Taf06}
    Tafalla M., Santiago-\Garcia J., Myers P. C., Caselli P., Walmsley C. M., 
    Crapsi A., 2006, A\&A, 455, 577

\bibitem[\protect\citeauthoryear{Turner}{2001}]{turner01} Turner 
B.~E., 2001, ApJS, 136, 579 

\bibitem[\protect\citeauthoryear{Turner, Pirogov, 
\& Minh}{1997}]{turner97} Turner B.~E., Pirogov L., Minh Y.~C., 1997, ApJ, 483, 235 

 \bibitem[\protect\citeauthoryear{Walmsley et al.}{1982}]{Walms82}
    Walmsley C. M., Churchwell E., Nash E., Fitzpatrick E., 1982, ApJ, 258, 75

\bibitem[\protect\citeauthoryear{Wong et al.}{2008}]{wong08} 
Wong T., et al., 2008, MNRAS, 386, 1069

\bibitem[{{Wu} \& {Evans}(2003)}]{wu03}
{Wu}, J. \& {Evans}, II, N.~J. 2003, ApJL, 592, L79

\bibitem[\protect\citeauthoryear{Wu et al.}{2005}]{Wu05}
    Wu Y., Zhu M., Wei Y., Xu D., Zhang Q., Fiege J. D., 2005, ApJ, 628, 57

\bibitem[{{Yun} {et~al.}(1999){Yun}, {Moreira}, {Afonso}, \& {Clemens}}]{yun99}
{Yun}, J.~L., {Moreira}, M.~C., {Afonso}, J.~M., \& {Clemens}, D.~P. 1999, AJ,
  118, 990

\bibitem[\protect\citeauthoryear{Zinchenko \& Pirogov}{1987}]{Zin87}
    Zinchenko I. I., Pirogov L. E., 1987, SvA, 31, 254
\end{thebibliography}
\end{document}